\documentclass[useAMS,usenatbib]{mn2e}

\usepackage{amssymb}
\usepackage[fleqn]{amsmath}
\usepackage{times}
\usepackage{graphicx}
\usepackage{epstopdf}
\usepackage{subfigure}
\usepackage{mathrsfs}
\usepackage{xcolor}

\usepackage[colorlinks=true,linkcolor=blue,anchorcolor=blue,citecolor=blue,urlcolor=blue]{hyperref}

\newcommand{\beq}{\begin{equation}}
\newcommand{\eeq}{\end{equation}}

\def \Mpch {\, \mathrm{Mpc}\,h^{-1} \,}

\newcommand{\bfr}{\mathbf{r}}
\newcommand{\bfx}{\mathbf{x}}

\def\gs{\mathrel{\lower0.6ex\hbox{$\buildrel {\textstyle >}\over{\scriptstyle \sim}$}}}
\def\ls{\mathrel{\lower0.6ex\hbox{$\buildrel {\textstyle <}\over{\scriptstyle \sim}$}}}
\newcommand{\simgt}{\lower.5ex\hbox{$\; \buildrel > \over \sim \;$}}
\newcommand{\simlt}{\lower.5ex\hbox{$\; \buildrel < \over \sim \;$}}

\newcommand{\aap}{A\&A}
\newcommand{\apj}{ApJ}
\newcommand{\apjl}{ApJ}
\newcommand{\apjs}{ApJS}

\newcommand{\jcap}{J. Cosmol. Astropart. Phys.}

\newcommand{\prd}{Phys. Rev. D}
\newcommand{\mnras}{MNRAS}

\def \nat {Nature}

\begin{document}

\title[$\sigma_8$ from lensing and clustering of clusters] {New constraints on $\sigma_8$ from a joint analysis of stacked gravitational lensing and clustering of galaxy clusters}

\author[Sereno, Veropalumbo, Marulli, et al.]{Mauro Sereno$^{1,2}$, Alfonso Veropalumbo$^{1}$, Federico Marulli$^{1,2,3}$, Giovanni Covone$^{4,5}$, \newauthor
Lauro Moscardini$^{1,2,3}$, Andrea Cimatti$^{1}$\\
  $^1$Dipartimento di Fisica e Astronomia, Alma Mater Studiorum -- Universit\`a di Bologna, Viale Berti Pichat 6/2, 40127 Bologna, Italia\\ 
  $^2$INAF, Osservatorio Astronomico di Bologna, via Ranzani 1, 40127 Bologna, Italia\\ 
  $^3$INFN, Sezione di Bologna, Viale Berti Pichat 6/2, 40127 Bologna, Italia\\ 
  $^4$Dipartimento di Fisica, Universit\`a di Napoli `Federico II', Via Cinthia, 80126 Napoli, Italia\\
  $^5$INFN, Sezione di Napoli, Compl. Univ. Monte S. Angelo, Via Cinthia, 80126 Napoli, Italia
  }

\maketitle

\begin{abstract}
The joint analysis of clustering and stacked gravitational lensing of galaxy clusters in large surveys can constrain the formation and evolution of structures and the cosmological parameters. On scales outside a few virial radii, the halo bias, $b$, is linear and the lensing signal is dominated by the correlated distribution of matter around galaxy clusters. We discuss a method to measure the power spectrum amplitude $\sigma_8$ and $b$ based on a minimal modelling. We considered a sample of $\sim 120000$ clusters photometrically selected from the Sloan Digital Sky Survey in the redshift range $0.1<z<0.6$. The auto-correlation was studied through the two-point function of a subsample of $\sim 70000$ clusters; the matter-halo correlation was derived from the weak lensing signal of the subsample of $\sim 1200$ clusters with Canada-France-Hawaii Lensing Survey data. We obtained a direct measurement of $b$, which increases with mass in agreement with predictions of the $\Lambda$CDM paradigm. Assuming $\Omega_\mathrm{M}=0.3$, we found $\sigma_8=0.78\pm0.16$. We used the same clusters for measuring both lensing and clustering and the estimate of $\sigma_8$ did require neither the mass-richness relation, nor the knowledge of the selection function, nor the modelling of $b$.  With an additional theoretical prior on the bias, we obtained  $\sigma_8=0.75\pm0.08$.
\end{abstract}

\begin{keywords}
galaxies: clusters: general -- gravitational lensing: weak -- cosmological parameters -- large-scale structure of Universe.
\end{keywords}

\section{Introduction}
\label{ss|intro}
Measurements of the large-scale structure growth can constrain the cosmological scenario and the formation and evolution of cosmic structures \citep[ and references therein]{euclid_ame+al13}. The study of growth as a function of time can determine the initial amplitude of matter fluctuations, the matter density, and the nature of dark energy. Furthermore, the scale dependence of structure growth can be used to constrain the neutrino mass \citep{vil+al14}.

A basic approach to study the structure growth exploits the simple hypothesis that galaxies trace the matter density field, whose correlation function is very rich in astrophysical and cosmological information and it is most easily predicted by the theory. Stacked gravitational lensing is the cross-correlation between foreground deflector positions and background galaxy shears. Weak gravitational lensing depends on the total matter density (including dark matter) via the deflection of light due to intervening matter along the line of sight, which both magnifies and distorts galaxy shapes. Stacked lensing can then be used to measure the galaxy-mass cross-correlation. On the other hand, galaxy clustering recovers the auto-correlation of galaxy positions.

Galaxies are biased tracers of the underlying mass distribution \citep{sh+to99,tin+al10,bha+al11}. This severely limits the constraining power of either galaxy clustering or stacked lensing on the matter power spectrum amplitude. Constraints from the two probes have to be combined to break degeneracies and to recover the matter correlation function \citep{bal+al10,og+ta11,man+al13,cac+al13,miy+al13,mor+al14}.

Theory and numerical simulations shows that the galaxy bias is extremely complicated to model: it is stochastic, it depends on galaxy properties such as luminosity, colour and/or morphological type, and it is scale dependent on small scales \citep[][ and references therein]{sh+to99,cac+al13, maru+al13}. The proper treatment of how galaxies populate dark matter haloes, assembly bias, and baryonic effects on the matter power spectrum on small scales requires a very accurate modelling. If the adopted scheme is too restrictive and fails to account for some important features, the constraining power on cosmological parameters is limited and the results can be severely biased.

Galaxy bias contains very valuable information regarding galaxy formation, mainly on small scales, but at the same time it is difficult to  model it properly. Complementary approaches have been proposed to deal with bias in joint clustering plus lensing analyses. At one extreme, methods can be optimised to study the bias. Physically motivated models based on the halo occupation distribution (HOD) have been considered to simultaneously solve for cosmology and galaxy bias \citep{yoo+al06,lea+al12,tin+al12,cac+al13}. Using the small scale lensing signals enhances the signal-to-noise ratio (SNR) and consequently reduces the statistical errors. However, problems connected to theory interpretation, arbitrary bias modelling, and observational uncertainties are more pronounced on small scales and they can cause additional systematic uncertainties which are difficult to ascertain.

At the other extreme, galaxy bias can be seen as a nuisance when attempting to determine cosmological parameters. The information from galaxy clustering and galaxy-galaxy lensing can then be  retained only above scales equal to a few times the typical dark matter halo virial radius, where the treatment of the bias is simplified \citep{bal+al10,yo+se12,man+al13}.

Most of the previous studies which combine clustering and lensing have focused on galactic scales. These studies can be optimised to estimate $\sigma_8$, i.e., the root mean square mass fluctuation amplitude in spheres of size 8$h^{-1}$Mpc. \citet{man+al13} recently constrained cosmology and galaxy bias using measurements of galaxy abundances, galaxy clustering and galaxy-galaxy lensing taken from the Sloan Digital Sky Survey (SDSS) data release 7. In the framework of the cold dark matter model with a cosmological constant ($\Lambda$CDM), they found $\sigma_8=0.76\pm0.06$. \citet{mor+al14} measured the clustering and abundance of the BOSS (Baryon Oscillation Spectroscopic Survey) galaxies from the SDSS-III (data release 11), and their galaxy-galaxy lensing signal with the CFHTLenS to find $\sigma_8=0.79\pm0.05$.

In the era of precision cosmology, the development of independent methods to measure cosmological parameters is crucial to test possible failures of the standard $\Lambda$CDM model. The tension between the lower values of $\sigma_8$ inferred from clusters counts \citep[ and references therein]{planck_2013_XX} and higher estimates from measurements of the primary Cosmic Microwave Background (CMB) temperature anisotropies \citep{planck_2013_XVI} may reflect either the need to extend the minimal $\Lambda$CDM model or some hidden systematics.

The \citet{planck_2013_XX} measured $\sigma_8=0.75\pm0.03$ and  $\Omega_\mathrm{M}=0.29\pm0.02$ using number counts as a function of redshift of 189 galaxy clusters from the Planck Sunyaev-Zel'dovich catalogue. However, the values of the cosmological parameters obtained from cluster abundance are degenerate with any systematic error in the assumed scaling relation between mass and the observed quantity. This problem can be solved in the context of joint experiments alike that considered in this paper. In fact, the combination of cluster observables (number counts and cluster-cluster correlation functions) and stacked weak lensing enables secure self-calibration of important systematic errors inherent in these measurements, including the source redshift uncertainty and the cluster mass-observable relation \citep{og+ta11}.

Analyses of the primary CMB temperature anisotropies have provided higher estimates of the power spectrum amplitude. The \citet{planck_2013_XVI} found $\sigma_8=0.83\pm0.02$ assuming a standard flat $\Lambda$CDM model. Estimates of cosmological parameters with CMB experiment are very accurate but highly degenerate, since only one source redshift can be observed. Results are then model dependent.

Measurements of the cosmic shear, i.e., the auto-correlation of galaxy shape distortions due to intervening matter along the line of sight, can constrain the amplitude and growth of matter fluctuations. Using non-linear models of the dark-matter power spectrum, \citet{kil+al13} estimated $\sigma_8= 0.84 \pm 0.03$ for a flat $\Lambda$CDM model with $\Omega_\mathrm{M}=0.3$ from 2D large-scale structure weak gravitational lensing in the CFHTLenS. This method is not affected by halo bias but since it relies on auto-correlations rather than shear cross-correlations, coherent additive errors in galaxy shapes (such as those induced by seeing or distortions in the telescope) may be difficult to remove from the analysis \citep{man+al13}. Moreover, intrinsic alignments with the local density field anti-correlate with the real gravitational shear and can contaminate cosmic shear measurements \citep{hi+se04}.

Here we propose a novel method based on the joint analysis of clustering and lensing of clusters of galaxies. The focus on clusters of galaxies is intended for a much simpler discussion. Clusters of galaxies trace the biggest collapsed structures and produce a very clean lensing signal. The stacked lensing technique has been highly successful in measuring the average masses of galaxy clusters down to the less massive haloes \citep{joh+al07,man+al08,cov+al14,for+al14b}. Furthermore, galaxy clusters are more strongly clustered than galaxies. Measurements of the two-point correlation function of galaxy clusters have already provided detections of the baryon acoustic oscillations (BAO) peak \citep{est+al09,hut10,hon+al12,ver+al14}.

The novelty of the method is that: \textit{i)} we track clusters of galaxies rather than galaxies; \textit{iii)} we consider the same clusters for both lensing and clustering; \textit{ii)} we determine bias and $\sigma_8$ based exclusively on the large-scale signal. Even though some of these elements were separately considered by previous papers, their combination makes for a new approach with minimal modelling.

The method strongly relies on using the same cluster population for both stacked lensing and clustering. If we correlate the positions of the lenses, the galaxy bias for the considered sample can be directly measured without any demanding theoretical modelling. Instead of being a systematic uncertainty, the information on the bias can be extracted to constrain structure formation and evolution. We relate the bias to the observed cluster population rather than trying to model the bias as a function of the halo mass, which would require the problematic calibration of the mass against the observable property the clusters were selected for \citep{se+et14,ser+al14b}. At the same time, $\sigma_8$ can be estimated without the knowledge of the selection function of the clusters.

The simultaneous analysis of stacked lensing and clustering is further simplified by keeping only the information well beyond the virial radius. Even at large scales, the proper treatment of the connection between galaxies and dark matter requires the modelling of the halo occupation statistics as a function of galaxy luminosity through the conditional luminosity function, combined with the halo model, which describes the non-linear matter field in terms of its halo building blocks \citep{man+al13}. The modelling of halo bias from very massive haloes is instead much easier to perform on a theoretical ground \citep{tin+al10}.

We focused on the determination of the amplitude of the power spectrum, $\sigma_8$, in a reference flat $\Lambda$CDM model with matter density parameter $\Omega_\mathrm{M}=0.3$, baryonic density parameter $\Omega_\mathrm{B}=0.04$, spectral index $n_\mathrm{s}=1$ and Hubble constant $H_0=100~h\mathrm{km~s}^{-1}\mathrm{Mpc}^{-1}$ \citep{planck_2013_XVI}. When necessary, we assumed $h=0.7$.

The structure of the paper is as follows. In Section~\ref{sec_over} we present the basics of how the combination of galaxy clustering and stacked lensing can determine the amplitude of the power spectrum and the halo bias. In Section~\ref{sec_data}, we introduce the cluster catalog and the data-sets. Sections~\ref{sec_lens} and~\ref{sec_clus} detail how we performed the analysis of stacked lensing and clustering, respectively. The joint analysis and the cosmological constraints are presented in Section~\ref{sec_join}. Section~\ref{sec_fore} forecasts the performance of the method with Euclid data. Some future developments are prospected in Section~\ref{sec_pros}. Section~\ref{sec_conc} is devoted to the final considerations.

\section{Overview}
\label{sec_over}

Clustered haloes are biased tracers of the underlying mass distribution \citep{kai84,tin+al10}. Very massive and luminous haloes are preferentially found in regions of the Universe with above average density and are biased with respect to the mean dark matter distribution. 

Auto-correlation functions between either matter or halo density fields depend on the halo bias in different ways. In principle, we can break degeneracies with proper combinations of correlation functions and we can infer at the same time the cosmological parameters and the halo bias. 

The matter auto-correlation function is
\beq
\xi_\mathrm{mm}(\bfr)=\langle \delta_\mathrm{m}(\bfx)  \delta^*_\mathrm{m}(\bfx+\bfr) \rangle \, ,
\eeq
where $\delta_\mathrm{m}$ is the matter density contrast. The analogous auto-correlation function for the halo density field is $\xi_\mathrm{hh}(\bfr)$, which is related to the matter statistics through the halo bias $b$ as
\beq
\xi_\mathrm{hh}(r) = b^2(r) \xi_\mathrm{mm}(r) .
\eeq
On the large scales probed by clusters of galaxies, the cross-correlation coefficient between the matter and halo fluctuations is one and the bias is linear, i.e., $b(r) =$ constant \citep{man+al13}. For a given cosmological model, $b$ depends on the mass and the redshift of the haloes hosting the galaxies through the peak height \citep{sh+to99,tin+al10,bha+al11}. 

Finally, the cross-correlation function is
\beq
\xi_\mathrm{hm}(\bfr)=\langle \delta_\mathrm{g}(\bfx)  \delta^*_\mathrm{m}(\bfx+\bfr) \rangle.
\eeq
On the observational side, $\xi_\mathrm{hh}$ and $\xi_\mathrm{hm}$ can be measured through clustering and stacked lensing, respectively. If we focus on the bias and $\sigma_8$, we can single out simple
proportionality factors,
\beq
\xi_\mathrm{mm}\propto \sigma_8^2 ,\  \ \xi_\mathrm{hm}\propto b~\sigma_8^2, \ \ \xi_\mathrm{hh}\propto b^2 \sigma_8^2.
\eeq
In the regime where the bias is linear, $b$ and $\sigma_8$ can then be determined as
\beq
b \propto \frac{\xi_\mathrm{hh}}{\xi_\mathrm{hm}}, \  \sigma_8 \propto \sqrt{\frac{\xi_\mathrm{hm}^2}{\xi_\mathrm{hh}}}.
\eeq

\section{Data}
\label{sec_data}

Our reference catalogue is the sample of 132684 optically selected clusters of galaxies identified from the Sloan Digital Sky Survey III (SDSS-III, data release 8) by \citet[ WHL]{wen+al12}\footnote{The latest version of the WHL catalogue {\it cluster\_dr9sz.dat} is publicly available at \url{http://zmtt.bao.ac.cn/galaxy\_clusters/}.}. Over-densities of galaxies around the brightest cluster galaxies (BCGs) were identified through their photometric redshifts. The optical richness is defined as $R_{L_*}= \tilde{L}_{200} / L_*$, where $\tilde{L}_{200}$ is the total $r$-band luminosity within an empirically determined radius $\tilde{r}_{200}$ and $L_*$ is the evolved characteristic galaxy luminosity. $N_{200}$ is the number of member candidates within $\tilde{r}_{200}$. The subscript 200 denotes the radius $r_{200}$ and related quantities measured in a sphere whose mean density is 200 times the critical density at the halo redshift. The cluster photometric redshift reported in the catalogue is the median value of the photometric redshifts of the galaxy members.

A candidate cluster makes the final cut if $R_{L_*}\ge 12$ and $N_{200}\ge 8$. These thresholds corresponds to a mass of $\sim 0.6 \times 10^{14} M_{\odot}$ \citep{wen+al12,cov+al14}. The detection rate above this mass is estimated at $\sim 75\%$. The false detection rate is less than $6$ per cent for the whole sample. The catalogue spans a redshift range $0.05<z<0.8$.

To optimise the lensing signal, we considered a redshift limited subsample in the range $0.1<z<0.6$ (median redshift $z=0.365$), wherein $123822$ clusters lie (93.3 per cent of the whole sample). The upper redshift limit enabled us to perform a robust separation between the lensing and the background population \citep{cov+al14}.

For the computation of the two-point correlation function, we considered a subsample of $69527$ galaxy clusters, selected on a contiguous area of $\sim 9000$ deg$^2$ in the Northern Galactic Cap obtained including all the SDSS stripes between $10$  and $37$.  This selection is used to ease the reconstruction of the visibility mask. Nevertheless, its impact on the final results is  negligible, considering the uncertainties in the measurements.

For the lensing sample, we considered the 1176 clusters centred in the four fields of the Canada-France-Hawaii Telescope Lensing Survey \citep[CFHTLenS,][]{hey+al12}, covering about 154 square degrees in optical $ugriz$ bands. The public archive\footnote{\url{http://www.cfht.hawaii.edu/Science/CFHLS}.} provides weak lensing data processed with THELI \citep{erb+al13}, shear measurements with {\it lensfit} \citep{mil+al13}, and photometric redshift measurements with accuracy $\sim 0.04(1+z)$ and a catastrophic outlier rate of about 4 per cent \citep{hil+al12,ben+al13}. Full details on the shear measurements can be found in \citet{hey+al12}. Since we took all clusters in the CFHTLenS fields without any further restriction, the lensing clusters we considered are a small but unbiased subsample of the total catalog. This was verified with a Kolmogorov-Smirnov test. The optical richness (redshift) distributions are compatible with a probability of 49.0 (39.4) per cent. The main properties of the cluster samples are reported in Table~\ref{tab_bias_sigma8}.

The CFHTLenS is at the same time much deeper and much smaller than the SDSS. As far as the signal-to-noise ratio of the stacked haloes is concerned, these effects counterbalance each other and lensing results are comparable \citep{joh+al07,man+al08,man+al13,bri+al13,hud+al13,cov+al14,for+al14, for+al14b,vel+al14,ogu14}. A similar choice of data-sets for a joint analysis of galaxy clustering and galaxy-galaxy lensing was recently made by \citet{mor+al14}.

Since the cluster catalogue and the shape measurements we considered are extracted from completely different data-sets, the SDSS and CFHTLenS data respectively, we are assured that the distribution of lens galaxies is uncorrelated with residual systematics in the shape measurements \citep{miy+al13}.

\section {Weak lensing}
\label{sec_lens}

The connection between galaxies and matter can be probed with gravitational lensing. In this section, we review the methods used to extract the lensing signal from the shape measurements, to estimate the observational uncertainties, and to constrain the cosmological parameters.

\subsection{Basics}
The so-called shear-cluster correlation (or stacked lensing) is the cross-correlation between the cluster distribution and the shapes of source galaxies. The main observable quantity for weak lensing is the tangential shear distortion $\gamma_+$ of the shapes of background galaxies. It is related to the projected surface density $\Sigma (R)$ around lenses \citep{man+al13}, 
\beq
\label{eq_Sigma}
\Sigma(R) = \bar{\rho}_\mathrm{m} \int \left[ 1+ \xi_\mathrm{hm}
  (\sqrt{R^2+\Pi^2}) \right] d \Pi, 
\eeq
via
\beq
\label{eq_Delta_Sigma}
\Delta \Sigma_+(R) = \gamma_+ \Sigma_\mathrm{cr} =
\bar{\Sigma}(<R)-\Sigma(R).  
\eeq 
In the equations above, $\bar{\rho}_\mathrm{m}$ is the mean mass density at $z$, $\Pi$ is the line of sight separation measured from the lens, $\bar{\Sigma}(<R)$ is the average lens matter density within the projected distance $R$, and $\Sigma_\mathrm{cr}$ is the critical surface density for lensing. For a single source redshift 
\beq 
\Sigma_\mathrm{cr}=\frac{c^2}{4\pi G} \frac{D_\mathrm{s}}{D_\mathrm{d} D_\mathrm{ds}}, 
\eeq where $c$ is the speed of light in the vacuum, $G$ is the gravitational constant, and $D_\mathrm{d}$, $D_\mathrm{s}$ and $D_\mathrm{ds}$ are the angular diameter distances to the lens, to the source, and from the lens to the source, respectively.

\subsection{Shear profile modelling}

Stacked lensing by galaxy clusters is described in terms of three main terms. The treatment is simplified with respect to galaxy-galaxy lensing, when central haloes have to be differentiated from satellites and related additional terms contribute to the total shear profile. 

Our treatment follows \citet{cov+al14}. The dominant contribution up to $\sim 1~\mathrm{Mpc}/h$ comes from the central haloes, $\Delta \Sigma_\mathrm{BMO}$. We modelled this term as a smoothly truncated Navarro-Frenk-White (NFW) density profile \citep[ BMO]{bal+al09},
\beq 
\rho_\mathrm{BMO} = \frac{\rho_\mathrm{s}}{\frac{r}{r_\mathrm{s}} (1 +  \frac{r}{r_\mathrm{s}})^2} \, \left(\frac{r_\mathrm{t}^2}{r^2 +  r_\mathrm{t}^2} \right)^2,
\eeq 
where $r_\mathrm{s}$ is the inner scale length, $r_\mathrm{t}$ is the truncation radius, and $\rho_\mathrm{s}$ is the characteristic density. When fitting the shear profiles up to very large radii (ten times the virial radius and beyond), the truncated NFW model gives less biased estimates of mass and concentration with respect to the original NFW profile \citep{og+ha11}. The truncation removes the unphysical divergence of the total mass and better describes the transition between the cluster and the 2-halo term which occurs beyond the virial radius.

In the following, as a reference halo mass, we consider $M_{200}$, i.e., the mass in a sphere of radius $r_{200}$. The concentration is defined as $c_{200}=r_{200}/r_\mathrm{s}$. We set the truncation radius $r_\mathrm{t} = 3\,r_{200} $ \citep{og+ha11,cov+al14}.

The second contribution to the total profile comes from the off-centred clusters, $\Delta \Sigma_\mathrm{off}$. The BCG defining the cluster centre might be misidentified \citep{joh+al07}, which leads to underestimate $\Delta \Sigma(R)$ at small scales and to bias low the measurement of the concentration. Furthermore, even if properly identified, the BCG might not coincide with the matter centroid, but this effect is generally very small and negligible at the weak lensing scale \citep{geo+al12,zit+al12}. The azimuthally averaged profile of clusters which are displaced by a distance $R_\mathrm{s}$ in the lens plane is \citep{yan+al06}
\beq 
\Sigma(R|R_\mathrm{s})=\frac{1}{2\pi}\int_0^{2\pi} d\theta
\Sigma_\mathrm{cen}(\sqrt{R^2+R_\mathrm{s}^2+2R R_\mathrm{s}\cos
  \theta}) , 
\eeq 
where $\Sigma_\mathrm{cen}$ is the centred profile. We modelled the distribution of off-sets with an azimuthally symmetric Gaussian distribution \citep{joh+al07,hi+wh10},
\beq
P(R_\mathrm{s})=\frac{R_\mathrm{s}}{\sigma_\mathrm{s}^2}\exp
\left[ -\frac{1}{2}\left(  \frac{R_\mathrm{s}}{\sigma_\mathrm{s}}\right)^2 \right], 
\eeq where
$\sigma_\mathrm{s}$ is the scale length. The contribution of the off-centred haloes is then 
\beq \Sigma_\mathrm{off} (R)=\int d
R_\mathrm{s} P(R_\mathrm{s}) \Sigma(R|R_\mathrm{s}).  
\eeq
Typical scale lengths are of order of $\sigma_\mathrm{s} \sim 0.4~\mathrm{Mpc}/h$ \citep{joh+al07}. Miscentring mainly matters with regard to an unbiased determination of the cluster concentration. Its effect is minor on the scales where the correlated matter manifests through the 2-halo term. We assumed that a fraction $f_\mathrm{off}$ of the lenses is miscentred.

The third significant contribution to the total density profile is the 2-halo term, $\Delta \Sigma_\mathrm{2h}$, which describes the effects of the correlated matter distribution around the location of the galaxy cluster at scales $\ga 10$ Mpc. The 2-halo shear term around a single lens of mass $M$ at redshift $z$ for a single source redshift can be modelled as
\citep{og+ta11,og+ha11}
\begin{equation}
\gamma_{+, 2h} (\theta; M, z) = \int \frac{l d l}{2 \pi} J_2(l \theta) \frac { \bar{\rho}_\mathrm{m} (z) b(M; z)}{ (1+z)^3 \Sigma_\mathrm{cr}  D_\mathrm{d}^2(z)} P_\mathrm{m}(k_l; z),
\label{eq:gamma_t2}
\end{equation}
where $ \theta$ is the angular radius, $J_2$ is the second order Bessel function, and $k_l \equiv l / [ (1+z) D_\mathrm{d}(z) ]$. $P_\mathrm{m}(k_l; z)$ is the linear power spectrum, which was computed following \cite{ei+hu99}. Given the observational errors on the shear measurements, more accurate computations of $P_\mathrm{m}$ have a negligible impact on the final result.

The final profile for the total differential projected surface density is then
\beq
\Delta \Sigma_\mathrm{tot}= (1-f_\mathrm{off})\Delta \Sigma_\mathrm{BMO}+f_\mathrm{off}\Delta \Sigma_\mathrm{off}+\Delta \Sigma_\mathrm{2h}.
\eeq
The above model has five free parameters: the mass $M_{200}$ and the concentration $c_{200}$ of the clusters; the fraction of off-centred haloes, $f_\mathrm{off}$, and the scale length $\sigma_\mathrm{s}$ of the probability distribution of the off-sets; the product $b\, \sigma_8^2$, which determines the amplitude of the 2-halo term.

We considered only the cosmological information contained in the 2-halo term, whose signal is proportional to $\propto b\,\sigma_8^2$. Small-scale lensing has typically the best signal-to-noise ratio, but it may be subject to systematic uncertainties both in terms of theoretical interpretation and observational uncertainties \citep{man+al13}. We did not try to connect it directly to the halo bias and the cosmological background. We just modelled it in terms of a physically motivated model, i.e., the (truncated) NFW profile. In this basic approach, the parameters $M_{200}$ and $c_{200}$ of the main haloes, as well as the parameters $f_\mathrm{off}$ and $\sigma_\mathrm{s}$ describing the miscentred clusters, can be seen as nuisance parameters that we marginalise over to measure the amplitude of the 2-halo term.

However, gravitational lensing presents an exclusive feature: it provides a direct measurement of the halo mass without relying on any scaling relation. The estimate of $M_{200}$ can then be used to constrain the evolution of bias with the halo mass.

\subsection{Measured profiles}
\label{sec_lens_anal}

We measured the lensing signal behind the 1176 clusters of the WHL catalog centred in the four fields of the CFHTLenS. We extracted the shear profiles between 0.1 and 30~Mpc/$h$ in 25 radial annuli equally distanced in logarithmic space. The procedure is described in \citet{cov+al14}, which we refer to for details on shear calibration, selection of background galaxies, source redshift estimation and stacking.

Very briefly, the raw shear components in the CFHTLenS catalogue were corrected by applying a multiplicative and an additive parameter, empirically derived from the data. The background lensed galaxy population behind each galaxy cluster were selected by using a two colour selection in the $g$ and $r$ bands \citep{med+al10,ogu+al12}, which can safely single out galaxies at $z\ga0.7$. We determined the tangential and cross component of the shear for each cluster from the weighted ellipticity of the background sources. 

The clusters were finally stacked according to their optical richness. We adopted two binning schemes: either a single bin in optical richness comprising all clusters, which we will refer to in the following as our reference case, or four bins with comparable SNR ($12 \le R_{L^*} <16$, $16 \le R_{L^*} <21$, $21 \le R_{L^*} <30$ and $R_{L^*} \ge 30$), as already done in \citet{cov+al14}.

\subsection{Uncertainty covariance matrix}

\begin{figure}
      \resizebox{\hsize}{!}{\includegraphics{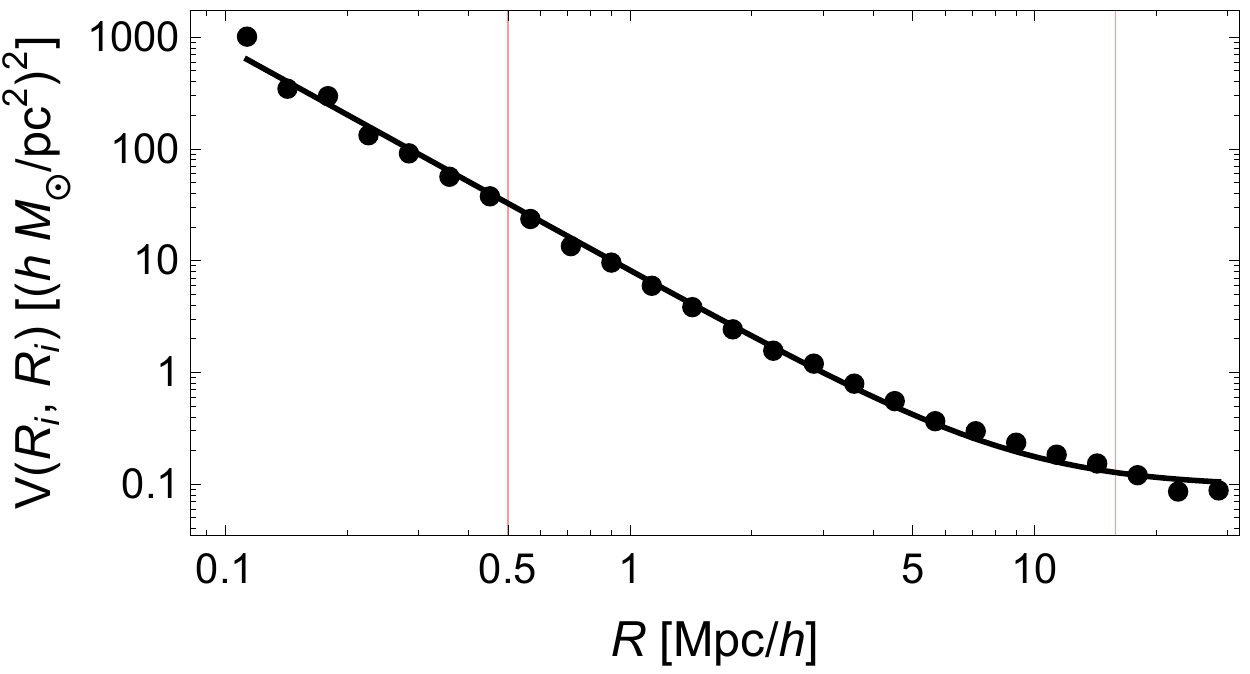}}
       \caption{Diagonal terms of the lensing uncertainty covariance matrix of the total sample ($R_{L^*}\ge 12$) as a function of the transverse separation $R$. The matrix elements were estimated with a bootstrap procedure. The full black line plots the fitted smoothing function. Vertical red lines delimit the radial range considered when fitting the shear profiles.}
	\label{fig_Delta_Covariance_diagonal}
\end{figure}

Due to stacking, shear observations at different radii are correlated. The effect is significant at radii larger than the typical lens angular separation \citep{man+al13}. We estimated the uncertainty covariance matrix with a bootstrap procedure with replacement. We resampled the clusters 10000 times. Covariance also accounts for the residual contribution from large scale projections, which is subdominant due to the large number of line of sights we stacked over.

The inverse of a noisy, unbiased estimator of the covariance matrix is not an unbiased estimator of the inverse covariance matrix \citep{har+al07,man+al13}. An unbiasing correction factor can be estimated under very restrictive statistical requirements, which are likely violated under usual observational conditions \citep{har+al07}. Furthermore, the correction is negligible if the number of lenses is significantly larger than the number of radial bins. We preferred not to apply any correction.

An alternative approach requires the smoothing of the covariance matrix to create a noiseless version \citep{man+al13}. The diagonal terms behave according to reproducible trends. The shape noise is the dominant source of variance at small radii. It scales like $R^{-2}$ for logarithmically spaced annular bins. However,  the total noise flattens at larger radii. There are two main reasons. First, when $R$ is significantly larger than the typical separation between lenses, annular bins include many of the same sources around nearby lenses and the shape noise can not decrease by adding more lenses \citep{man+al13}.

Secondly, when $R$ is comparable with the field of view of the camera, an imperfect correction of the optical distortion can cause a tangential or radial pattern of the point spread function (PSF) ellipticities in the edge of the field of view. This coherent PSF anisotropy can then cause a residual systematic error \citep{miy+al13}. The field of view of CFHT/MegaCam is $1\deg$ large, which corresponds to $\sim12.9~\mathrm{Mpc}/h$ at the median redshift of the lens sample, $z\sim 0.37$.

Taking into account all the above sources of noise, the diagonal terms of the lensing uncertainty covariance matrix $\mathbfss{V}$ can be modelled as
\citep{man+al13} 
\beq \mathbfss{V}(R_i,R_i) =A R^{-2}\left[ 1+  (R/R_\mathrm{t})^2 \right], 
\eeq 
where $R_\mathrm{t}$ denotes the turn-around radius above which the shape noise is subdominant. A basic unweighted fit for the reference binning in optical richness ($R_{L^*} \ge 12$) gives $R_\mathrm{t} \sim 9.3~\mathrm{Mpc}/h$ (see Fig.~\ref{fig_Delta_Covariance_diagonal}).

The smoothing of the non-diagonal terms of the covariance matrix is more problematic. The two main sources of correlation are the cosmic variance (which is subdominant given the large number of line of sights in our sample) and the correlated shape noise due to the large $R$ compared to the separation between lenses \citep{man+al13}. These effects are difficult to model. Furthermore, a smoothing procedure might bias low the estimated correlation of the elements near the diagonal. We then preferred to use the noisy version of the covariance matrix. 

The uncertainty covariance matrix could be determined only when the number of clusters to stack was large enough. This is not the case for the less populated bins in optical richness. We then decided to use the full covariance matrix only for our reference case, whereas we took only the diagonal elements for the subsamples in optical richness in order to perform a coherent analysis when we looked for trends with richness/mass.

\subsection{Random catalogues}

\begin{figure}
      \resizebox{\hsize}{!}{\includegraphics{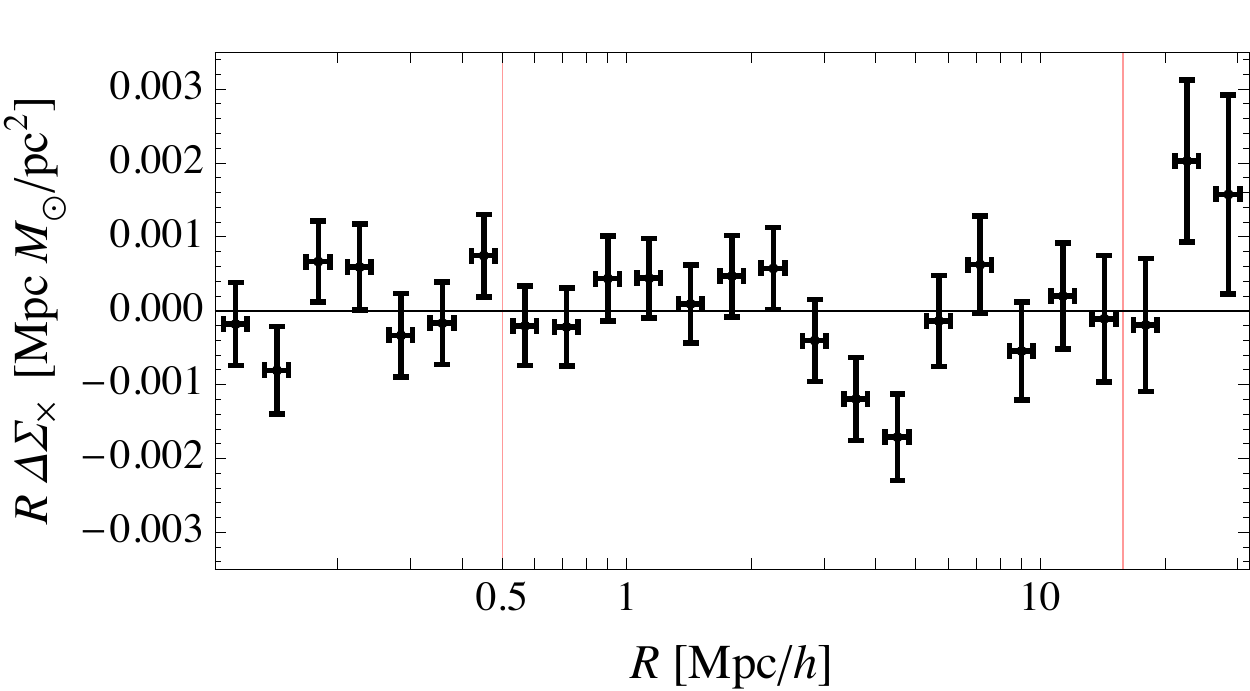}}
       \caption{The renormalized cross component of the differential shear profile of the full sample of lensing clusters ($R_{L^*}\ge 12$) after correction for the residual signal. Vertical red line delimit the radial range considered when fitting the shear profiles.}
	\label{fig_Delta_Sigma_X}
\end{figure}

Residual systematic effects affecting the stacked shear profiles may come either from stacking over annuli which are largely incomplete due to the limited field of view or from other source of errors. These additional systematics were estimated by extracting the signal around random points with the same procedure used for the cluster lenses. We built a catalogue of 5046 random lenses with the same redshift and spatial distribution of the galaxy clusters. We realised 10000 bootstrap resamplings with replacement of the catalog. The signal from the random pointings is consistent with zero up to $\sim 5~\mathrm{Mpc}/h$. The spurious signal at larger radii is likely due to residual systematics in the shear measurements at the edges of detector \citep{miy+al13}.

The shear profiles of the stacked clusters can be corrected for these residual systematics by subtracting the stacked signal estimated from the random catalog. This correction rests on the assumption that the distribution of lenses is uncorrelated with residual systematics in the shape measurements \citep{miy+al13}, which holds in our analysis because the cluster catalogue and the shape measurements are taken from completely different data-sets.

After correction for the spurious signal, the cross component of the shear profile, $\Delta \Sigma_\times =\Sigma_\mathrm{cr}\gamma_\times$, is consistent with zero at nearly all radii. This confirms that the main systematics have been eliminated. The radial profile of $\Delta \Sigma_\times$ for the full sample of clusters is plotted in Fig.~\ref{fig_Delta_Sigma_X}. Most of the points are within 1-$\sigma$ from the null value which might indicate somewhat over-estimated uncertainties. $\Delta \Sigma_\times$ increases at $R \ga 20~\mathrm{Mpc}/h$ but the deviation is not statistically significant.

\subsection{Constraints}
\label{sec_lens_cons}

\begin{figure}
      \resizebox{\hsize}{!}{\includegraphics{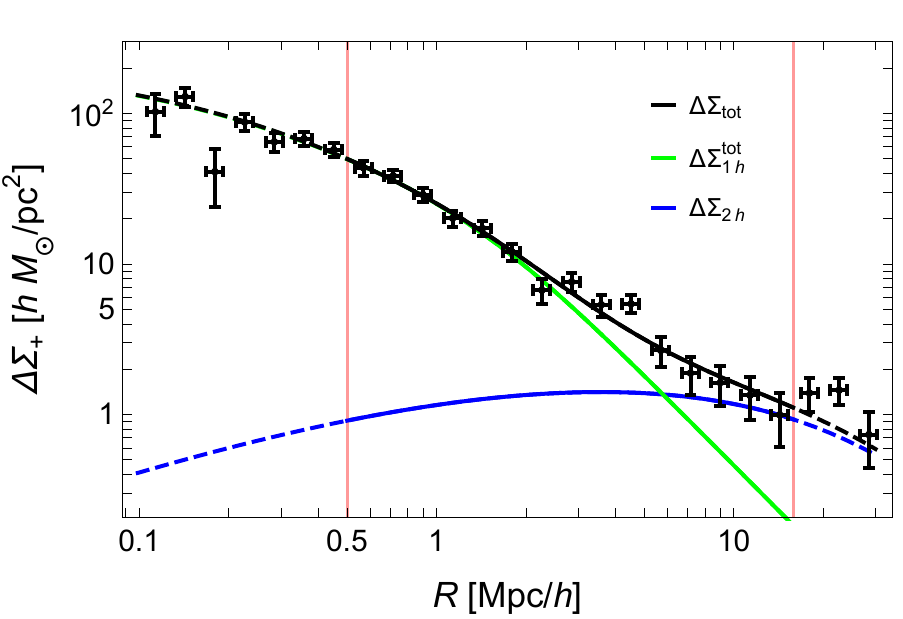}}
       \caption{Stacked differential surface density $\Delta \Sigma_+$ after correction for the residual signal as a function of radius for clusters with optical richness $R_{L^*}\ge 12$. Black points are our measurements. The vertical error bars show the square root of the diagonal values of the covariance matrix. The horizontal lines are the weighted standard deviations of the distribution of radial distances in the annulus. The green line plots the contribution from the galaxy cluster haloes (i.e., the sum of lensing contributions from centred and offset lenses); the blue line describes the 2-halo term; the black line is the overall fitted radial profile. Vertical red lines delimit the radial range considered when fitting the shear profiles. Dashed lines are extrapolations based on the best fit model, which was determined in the radial range $0.5<R<15.8\,\mathrm{Mpc/}h$. }
	\label{fig_Delta_Sigma_t}
\end{figure}

\begin{table}
\caption{Product $b\ \sigma_8^2$ (col.~2) for the reference sample ($R_{L^*}\ge12$) as determined with the stacked lensing analysis under different assumptions. In col.~1, we reported the difference in the fitting procedure with respect to the `reference' case. All else being equal, we considered: different priors on the truncation radius $r_\mathrm{t}$ of the BMO model of the central haloes; a standard NFW model; a uncertainty covariance matrix with null off-diagonal elements; no miscentred haloes; different thresholds for either the maximum ($r_\mathrm{max}$) or the minimum ($r_\mathrm{min}$) fitting radius.}
\label{tab_bias_sigma8Sq}
\centering
\begin{tabular}[c]{l  r@{$\,\pm\,$}l }
        \hline
	Assumption	& \multicolumn{2}{c}{$b~\sigma_8^2$}   \\
	\hline
	reference							&	1.56		&	0.35	\\
	$r_\mathrm{t}=4\ r_{200}$ 			&	1.52		&	0.34	\\	
	$r_\mathrm{t}=2\ r_{200}$			&	1.63		&	0.34	\\
	NFW	 							&	0.96		&	0.42	\\
	diagonal covariance					&	1.54		&	0.28	\\
	$f_\mathrm{off}=1$					&	1.58		&	0.33	\\
	$r_\mathrm{max}=20~\mathrm{Mpc}/h$	&	1.73		&	0.32	\\
	$r_\mathrm{max}=30~\mathrm{Mpc}/h$	&	1.86		&	0.29	\\
	$r_\mathrm{min}=0.1~\mathrm{Mpc}/h$	&	1.55		&	0.33	\\
	$r_\mathrm{min}=0.2~\mathrm{Mpc}/h$	&	1.55		&	0.34	\\
	\hline
	\end{tabular}
\end{table}

The corrected excess surface density for the total sample is plotted in Fig.~\ref{fig_Delta_Sigma_t}. The signal is detected with high significance ($\mathrm{SNR}\simeq26.1$) over the full radial range. Stacked profiles for subsamples in optical richness can be found in \citet{cov+al14}. Based on the analysis of the turn-around radius in the diagonal elements of the covariance matrix and the features of the shear profile $\Delta \Sigma_\mathrm{\times}$, we limited our analysis to a maximum radius of 15.8~Mpc/$h$.

The choice of the lower limit for the radial range comes from a compromise between minimising the systematic errors due to contamination of cluster member galaxies and non-linear effects, and minimising the statistical errors \citep{man+al08}. We considered a minimum radius of 0.5~Mpc/$h$. Since to measure $\sigma_8$ we used only the information in the 2-halo term, which is dominant at very large radii, our final constraints are affected in a very negligible way by the choice of the lower fitting radius, which mainly impacts the determination of the effective concentration of the stacked clusters.

Radial fits were then performed between 0.5 and 15.8~Mpc/$h$ (15 equally spaced points in logarithmic scale). The statistical analysis was based on a $\chi^2$ function,
\begin{equation}
\chi_\mathrm{WL}^2 = \sum_{i,j} \left[ \Delta
  \Sigma_{\mathrm{obs},i} -\Delta \Sigma_{\mathrm{th},i}
  \right]\mathbfss{V}^{-1}_{ij} \left[ \Delta
  \Sigma_{\mathrm{obs},j}-\Delta \Sigma_{\mathrm{th},j}\right],
\end{equation}
where $\mathbfss{V}^{-1}$ is the inverse of the uncertainty covariance matrix, $\Delta \Sigma_{\mathrm{obs},i}$ is the observed excess density at radius $R_i$, and $\Delta \Sigma_\mathrm{th}$ is the theoretical prediction of the halo model. We adopted uniform priors for the fitting parameters and sampled the posterior probability with four Monte Carlo Markov chains. Results are summarised in Tables~\ref{tab_bias_sigma8Sq} and \ref{tab_bias_sigma8}. 

For the reference sample, the product $b\ \sigma_8^2$ was recovered with an accuracy of $\sim 20$ per cent. This estimate is stable against variation in the fitting procedure, which reassure us about the proper treatment of systematics (see Table~\ref{tab_bias_sigma8Sq}). As far as we consider a truncated model for the lenses, the final results on $b\ \sigma_8^2$ are nearly independent of the specific modelling. The variations in the estimated $b\ \sigma_8^2$ due to different assumptions on the truncation radius of the $\Sigma_\mathrm{BMO}$ profile are not significant and they are much smaller than the statistical uncertainty. The case of the divergent NFW model is different. This model is unphysical at scales well beyond the virial radius and it would severely under-estimate the contribution of the 2-halo term. 

We checked that the results of our analysis are very similar whether using the full covariance matrix or only the diagonal terms. Neglecting the covariance in the shear measurements of near radial bins does not affect the central estimate. On the other hand, the uncertainties on the fitting parameters are slightly smaller. This agreement further supports our choice of using the (noisy) covariance matrix without any correction.

As expected, a different minimum radius in the fitting procedure has no effect on the estimate of the bias, which only depends on the signal at scales $\ga10~\mathrm{Mpc}/h$. The inclusion of small scales affects nevertheless the estimate of the mass and of the concentration of the central halo. At scales $R \sim 0.1~\mathrm{Mpc}/h$ a proper modelling of the lens requires the treatment of the BCG and of the baryonic component. Similar considerations hold for the treatment of the miscentred halos too. The fraction of halos with off-sets has a negligible impact on the estimate of the bias. 

Finally, the inclusion of shear measurements at large radial scales not fully covered by the field of view can bias the results.

\section{Clustering}
\label{sec_clus}

This section provides a general description of the methods used in this work to measure the halo clustering, to estimate the observational uncertainties, and to constrain the linear bias and $\sigma_8$. We refer to \citet{ver+al14} for further details.

\subsection{The two-point correlation function}
\label{ss|2pcf}
To compute comoving distances from angular coordinates and redshifts, we assumed the fiducial cosmological parameters reported in \S\ref{ss|intro}.  We used the Landy-Szalay estimator \citep{landy1993} to measure the monopole of the two-point correlation function in redshift-space, 
\begin{equation}
\xi(s)  = \frac{1}{RR(s)}\times\left[  DD(s)\frac{n_\mathrm{r}^2}{n_\mathrm{d}^2}-2DR(s)\frac{n_\mathrm{r}}{n_\mathrm{d}} +RR(s) \right] \, ,
\label{e|lsext}
\end{equation}
where $DD(s)$, $DR(s)$ and $RR(s)$ are the number of data-data, data-random and random-random pairs within a comoving separation $s\pm\Delta s /2$, $\Delta s$ is the bin size, and $n_\mathrm{r}$ and $n_\mathrm{d}$ are the comoving number densities of the random and cluster sample, respectively.  The correlation function was measured in the range $10<r/(\Mpch)<40$.

\subsection{Modelling the redshift-space cluster clustering}
\label{ss|rscf}

Measuring distances from observed redshifts introduces {\em geometric} and {\em dynamic} distortions in the correlation function measurements, the former due to the assumption of a fiducial cosmology, not necessarily the true one, the latter caused by perturbations in the cosmological redshifts due to peculiar motions. Moreover, the precision of redshift measurements influences the estimate of line-of-sight distances \citep{maru+al12}.  As a first approximation, the relation between the observed redshift, $z_\mathrm{obs}$, and the cosmological one, $z_\mathrm{c}$, reads
\begin{equation}
z_\mathrm{obs}  = z_\mathrm{c} + \frac{v_{\parallel}}{c}(1+z_\mathrm{c}) \pm \sigma_z \, ,
\label{e|zobs}
\end{equation} 
where $v_{\parallel}$ is the velocity component along the line of sight of the peculiar velocity of the object, and $\sigma_z$ is the redshift error. We ignored geometric distortions, whose effect is negligible with respect to dynamic distortions and photo-$z$ errors \citep{maru+al12}, and we kept the cosmological parameters fixed in the fitting procedure.

The redshift-space 2D power spectrum can be modelled in polar coordinates as follows \citep[see e.g.][]{kaiser1987, peacock2001}:
\begin{equation}
P(k,\mu) = P_\mathrm{DM}(k) (b+f\mu^2)^2 \exp(-k^2\mu^2\sigma^2) \, ,
\label{e|pkmu}
\end{equation}
where $k=\sqrt{k_{\perp}^2+k_{\parallel}^2}$, $k_{\perp}$ and $k_{\parallel}$ are the wave-vector components perpendicular and parallel to the line of sight, respectively, $\mu = k_{\parallel}/k$, $P_\mathrm{DM}(k)$ is the dark matter power spectrum, $b$ is the linear bias factor, $f$ is the growth rate of density fluctuations, and $\sigma$ is the displacement along the line of sight due to random perturbations of cosmological redshifts. Assuming standard gravity, we approximated the growth rate as $f\simeq\Omega_\mathrm{M}(z)^{\gamma}$, with $\gamma=0.545$. The $f\mu^2$ term parametrizes the coherent motions due to large-scale structures, enhancing the clustering signal on all scales.  The exponential cut-off term describes the random perturbations of the redshifts caused by both non-linear stochastic motions and redshift errors. It washes out the signal over a typical scale $k\sim 1/\sigma$, thus causing a scale-dependent effect.

To derive the monopole of the correlation function, we integrated Eq.~(\ref{e|pkmu}) over the angle $\mu$, and then Fourier anti-transformed. The solution can be written as follows
\begin{equation}
\xi(s) = b^2 \xi'(s) + b \xi''(s) + \xi'''(s) \, .
\label{e|genxi}
\end{equation}
The main term, $\xi'$, is the Fourier anti-transform of the monopole
$P'(k)$:
\begin{equation}
P'(k) = P_\mathrm{DM}(k) \frac{\sqrt{\pi}}{2 k \sigma} \mathrm{erf}(k\sigma) \, ,
\end{equation}
that corresponds to the model given by Eq.~(\ref{e|pkmu}) when neglecting the dynamic distortion term. The $\xi''$ and $\xi'''$ terms are the Fourier anti-transforms of 
\begin{equation}
P''(k) = \frac{f}{(k\sigma)^3} P_\mathrm{DM}(k) \left[
\frac{\sqrt{\pi}}{2}\mathrm{erf}(k\sigma) -
k\sigma\exp(-k^2\sigma^2)\right]\, ,
\end{equation}
and 
\begin{multline}
P'''(k) =
\frac{f^2}{(k\sigma)^5}P_\mathrm{DM}(k) \left\{ \frac{3\sqrt{\pi}}{8}\mathrm{erf}(k\sigma) \right. \\ \left. 
- \frac{k\sigma}{4}\left[2(k\sigma)^2+3\right]\exp(-k^2\sigma^2)
\right\} \, ,
\end{multline}
respectively.

In our case, photo-$z$ errors perturb the most the distance measurements along the line of sight. Indeed, as verified in a precedent study \citep{ver+al14}, the BCGs identified as cluster centres are close to the minimum of the cluster potential wells. Therefore their small-scale random motions are negligible with respect to photo-$z$ errors, and the effect of non-linear peculiar velocities at small scales, the so-called {\em fingers of God} effect, can be safely neglected.  The cut-off scale in Eq.~(\ref{e|pkmu}) can thus be written as
\begin{equation}
\sigma = \frac{c\sigma_z}{H(z)} \, , 
\label{e|sigma}
\end{equation} 
where $H(z)$ is the Hubble function computed at the median redshift of the sample, and $\sigma_z$ is the typical photo-$z$ error.

\subsection{Photo-$z$ errors}
\label{ss|sze}

\begin{figure}
\includegraphics[width=\linewidth]{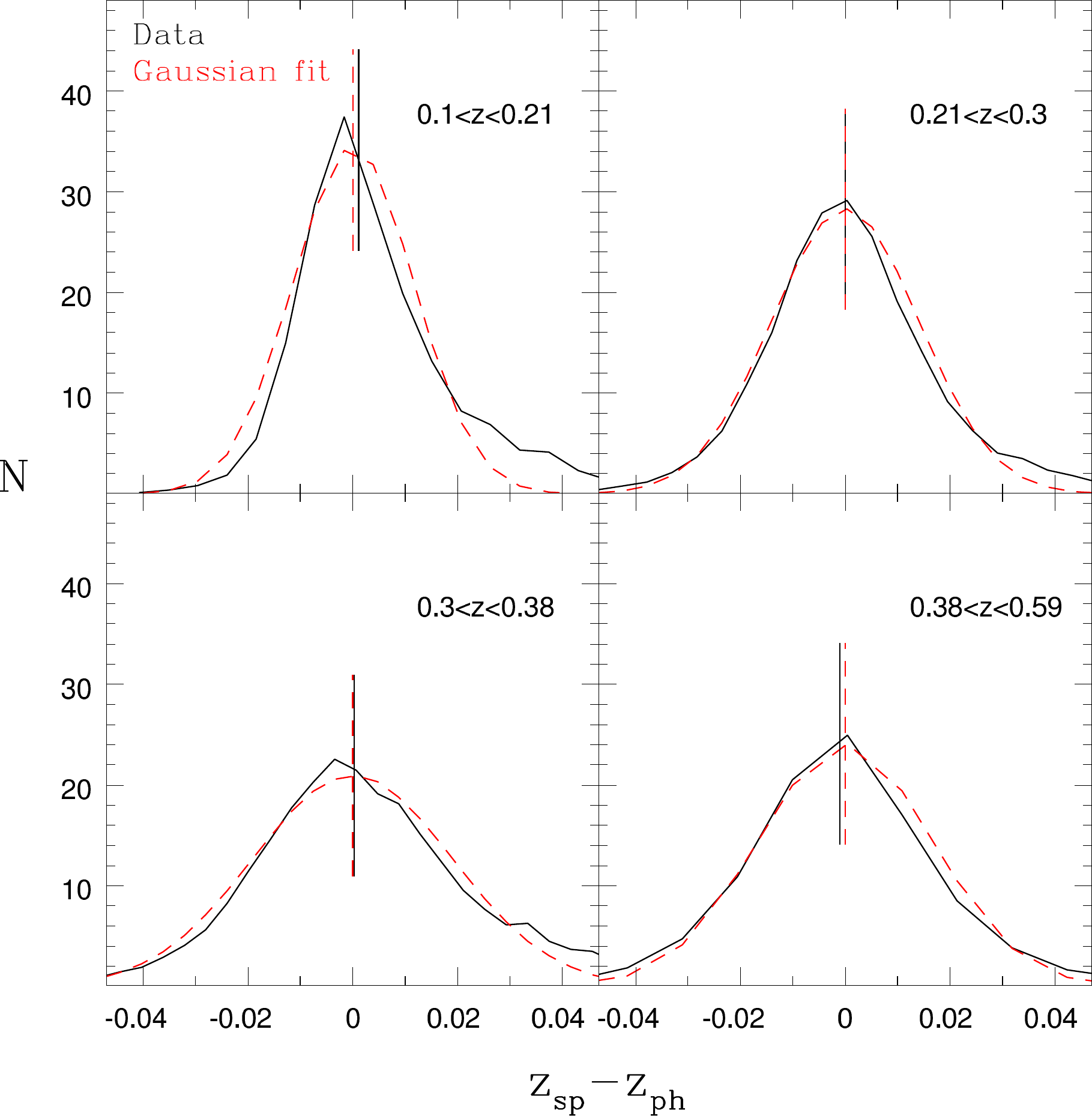} 
\caption{Distribution of $\Delta z(\equiv z_\mathrm{phot}-z_\mathrm{spec})$ in four redshift bins (black solid lines) and the associated Gaussian fit (red dashed lines). The limiting values of each redshift bin are indicated in the corresponding panels. The vertical lines correspond to the median values of the $\Delta z$ distributions (black) and to the mean of the Gaussian models (red dashed).}
\label{f|zg}
\end{figure}

\begin{figure}
	\centering
	\begin{tabular}{c}
	\includegraphics[width=0.95\linewidth]{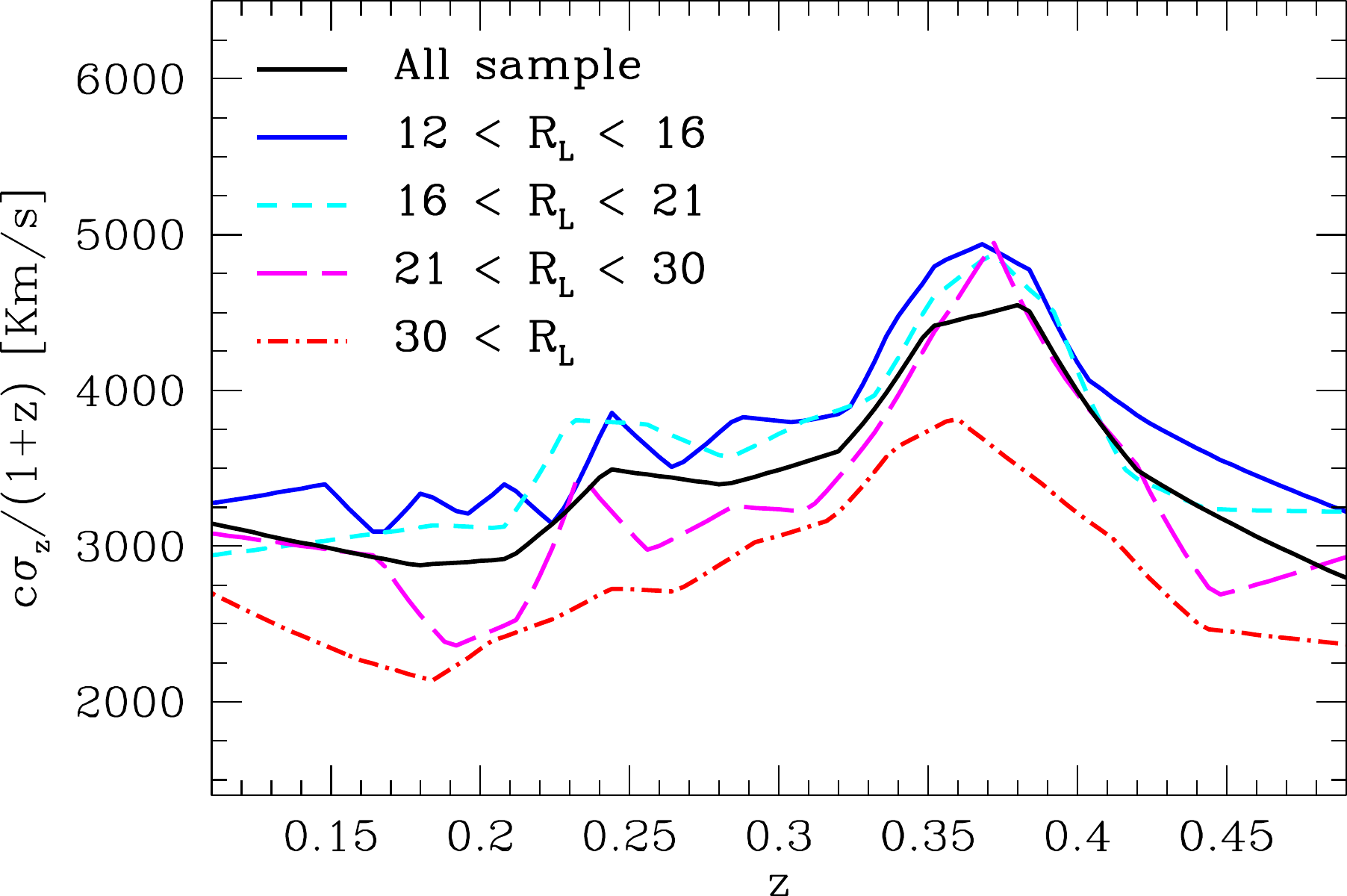} \\
	\\
	\includegraphics[width=0.95\linewidth]{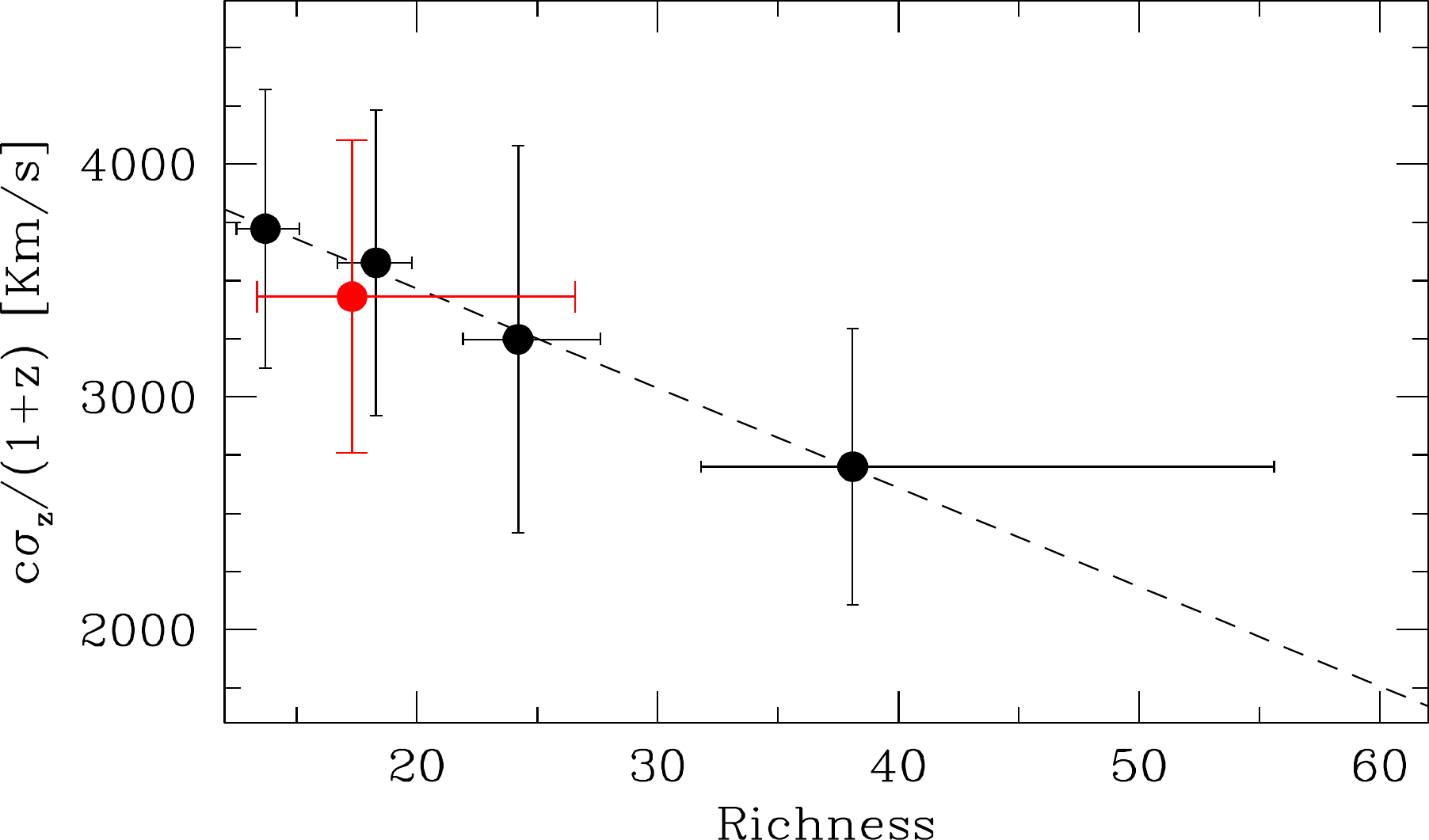} \\
\end{tabular}
	\caption{{\it Top panel:} photo-$z$ errors as a function of redshift for the whole sample (black solid line) and for the four richness bins reported in Table~\ref{tab_bias_sigma8}, as indicated by the labels. The photometric redshift is systematically better determined for high richness clusters. {\it Bottom panel:} photo-$z$ errors obtained following the procedure described in \S~\ref{ss|sze} for the whole sample (red point) and for the four richness bins.  The horizontal error bars span from the $16$ to the $84$ percentiles of the cluster richness distribution, while the vertical error bars are the standard deviations of the $\sigma_z$ redshift distribution. The solid line is a least-squares linear fit to the black points which highlights the decreasing behaviour as a function of richness.} 
\label{f|zv}
\end{figure}

Large photo-$z$ errors have a dramatic impact on the measured $\xi(s)$. The real-space clustering can be derived by projecting the correlation function along the line of sight \citep[e.g.][]{maru+al13, ver+al14}. However, this technique becomes quite ineffective for large redshift errors and small survey area, since it would be necessary to integrate the correlation function up to too large scales to fully correct for redshift errors. In this work we followed a different strategy, fitting directly the redshift-space clustering with a model that takes into account redshift uncertainties.\footnote{Due to targeting processes in SDSS, the redshift distributions of the photometric and spectroscopic redshift samples are different. Therefore, to combine clustering with lensing data, we were forced to measure the clustering of the photometric sample.}

To estimate the photo-$z$ errors, we used the spectroscopic data available for a subsample of the photometric catalogue. A spectroscopic redshift was assigned to a galaxy cluster if it was measured for its BCG. Spectroscopic redshifts of BCGs can be considered basically unaffected by peculiar velocities \citep{ver+al14}, and are measured with high precision. The resulting sample of clusters with both photometric and spectroscopic redshift measurements consists of $31338$ objects. 

We estimated the photo-$z$ errors, $\sigma_z$ by comparing spectroscopic and photometric cluster redshifts in different redshift and richness bins. As shown in Fig.~\ref{f|zg}, the $\Delta z (\equiv z_\mathrm{phot}-z_\mathrm{spec})$ distributions are well described by a Gaussian function at all redshifts. Specifically, Fig.~\ref{f|zg} shows the $\Delta z$ distributions (black solid lines) estimated in four illustrative redshift bins, $0.10\le z_\mathrm{phot}<0.21$, $0.21\le z_\mathrm{phot}<0.30$, $0.30 \le z_\mathrm{phot}<0.38$, and $0.38 \le z_\mathrm{phot}\le 0.59$. The red dashed lines show the best-fit Gaussian models, which faithfully reproduce the measured distributions. The photo-$z$ error, $\sigma_z$, could then be estimated as the standard deviation of the distribution.

To reconstruct the full $\Delta z$ distribution, we used $20$ redshift bins. Anyway, the estimate of the photo-$z$ errors resulted to be very weakly dependent on the specific redshift partition. In Fig.~\ref{f|zv}, top panel, we show the variation of $\sigma_z$ with redshift for the whole sample (black solid line) and for four richness bins (coloured lines) in the entire redshift range. The $z\sim 0.35$ peak is due to the shift of the $4000$~\AA ~break from the $g-r$ to $r-i$ colours \citep{ryk+14}. Our results agree with what found by \citet{wen+al12}. As it can be seen in the bottom panel of Fig.~\ref{f|zv}, thanks to the larger number of cluster members, the larger the richness, the smaller the photo-$z$ error. The estimated values of $\sigma_z$ are reported in Table \ref{tab_bias_sigma8}. Finally, to obtain a unique value of $\sigma_z$ to be used in Eq.~(\ref{e|sigma}), we averaged the standard deviations measured in the different redshift bins, weighting over the cluster redshift distribution.

\subsection{Random catalogues}
\label{ss|rs}

To construct the random catalogues, we reproduced separately the angular and redshift selection functions of the cluster sample. This method provides a fair approximation of the full distribution \citep{ver+al14}. The angular distribution of the random objects was obtained using the public software {\small MANGLE} \citep{swanson2008}. We first converted the measured angular positions to the SDSS coordinate system in order to reconstruct the footprint of the sample, that is eventually randomly filled. 

We associated redshifts to the random objects by drawing from the observed redshift distribution of the cluster samples. The latter was assessed grouping the data in $100$ redshift bins and applying a Gaussian convolution with a kernel three times larger than the bin size \citep{maru+al13}. Reducing the bin size has the effect of lowering the clustering signal along the line of sight. However, the impact of this effect is negligible considering the measurement uncertainties, as we verified. To minimise the impact of the shot noise at small scales, for each sample considered in this work, we generated a random catalogue $20$ times larger.

\subsection{Error estimates}
\label{ss|ee}

We used the {\em jackknife} re-sampling technique to estimate the covariance matrix of the correlation function measurements:
\begin{equation}
\mathbfss{C}_{ij}  = \frac{N_\mathrm{sub}-1}{N_\mathrm{sub}} \sum_{k=1}^N
(\xi_i^k-\bar{\xi}_i)(\xi_j^k-\bar{\xi}_j) \, ,
\label{eqcov}
\end{equation}
where $N_\mathrm{sub}$ is the number of resamplings, $\xi_i^k$ is the value of the correlation function in the $i$-th bin for the $k$-th subsample, and $\bar{\xi}_i$ is the mean value of the subsamples.

The {\em jackknife} subsamples are constructed following the SDSS geometry. We divided the original catalogue in fractions of SDSS stripes ($5$ regions in each of the $28$ considered SDSS stripes, $N_\mathrm{sub}=140$ subvolumes in total) and excluding recursively one of them. As tested, the number of subsamples is large enough to ensure convergence and stability of the covariance matrix estimate.  

We also tested our algorithms on the LasDamas mock samples \citep{mcbride2009}, finding that our error estimates are conservative with a typical over-estimation of $\sim$20 per cent \citep{ver+al14}.

\subsection{Constraints}
\label{ss|fr}

\begin{figure}
   \centering
   \includegraphics[width=\linewidth]{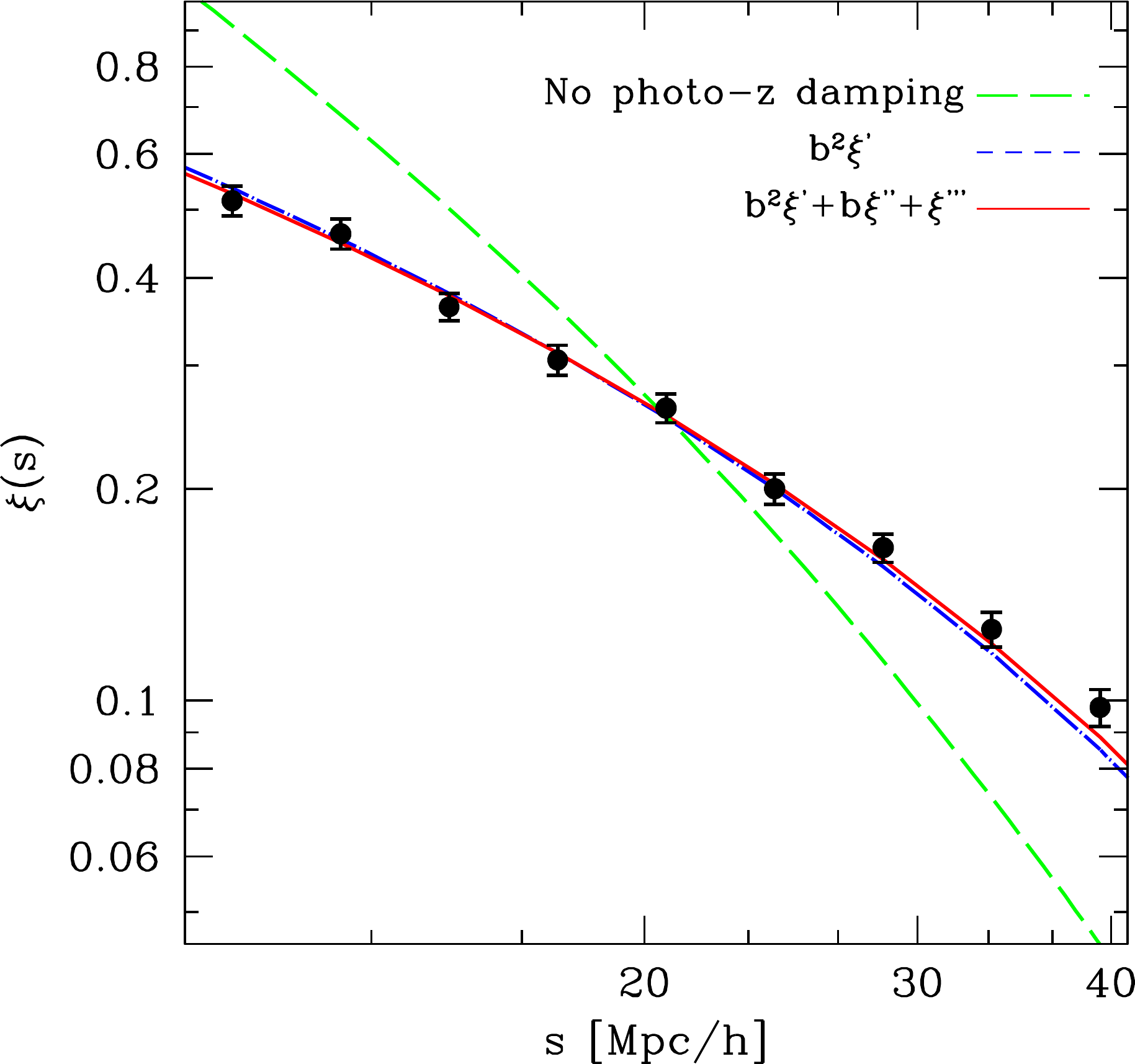}
   \caption{The redshift-space two-point correlation function of the whole cluster sample ($R_{L_*}\ge 12$, black dots), compared to best-fit models obtained with the full model in Eq.~(\ref{e|genxi}, red solid line), the dominant $\xi'$ term only (blue dashed line), and without the photo-$z$ damping term (green long-dashed line). The error bars show the square roots of the diagonal values of the covariance matrix.}
\label{f|xi}
\end{figure}

\begin{figure}
   \centering
   \includegraphics[width=\linewidth]{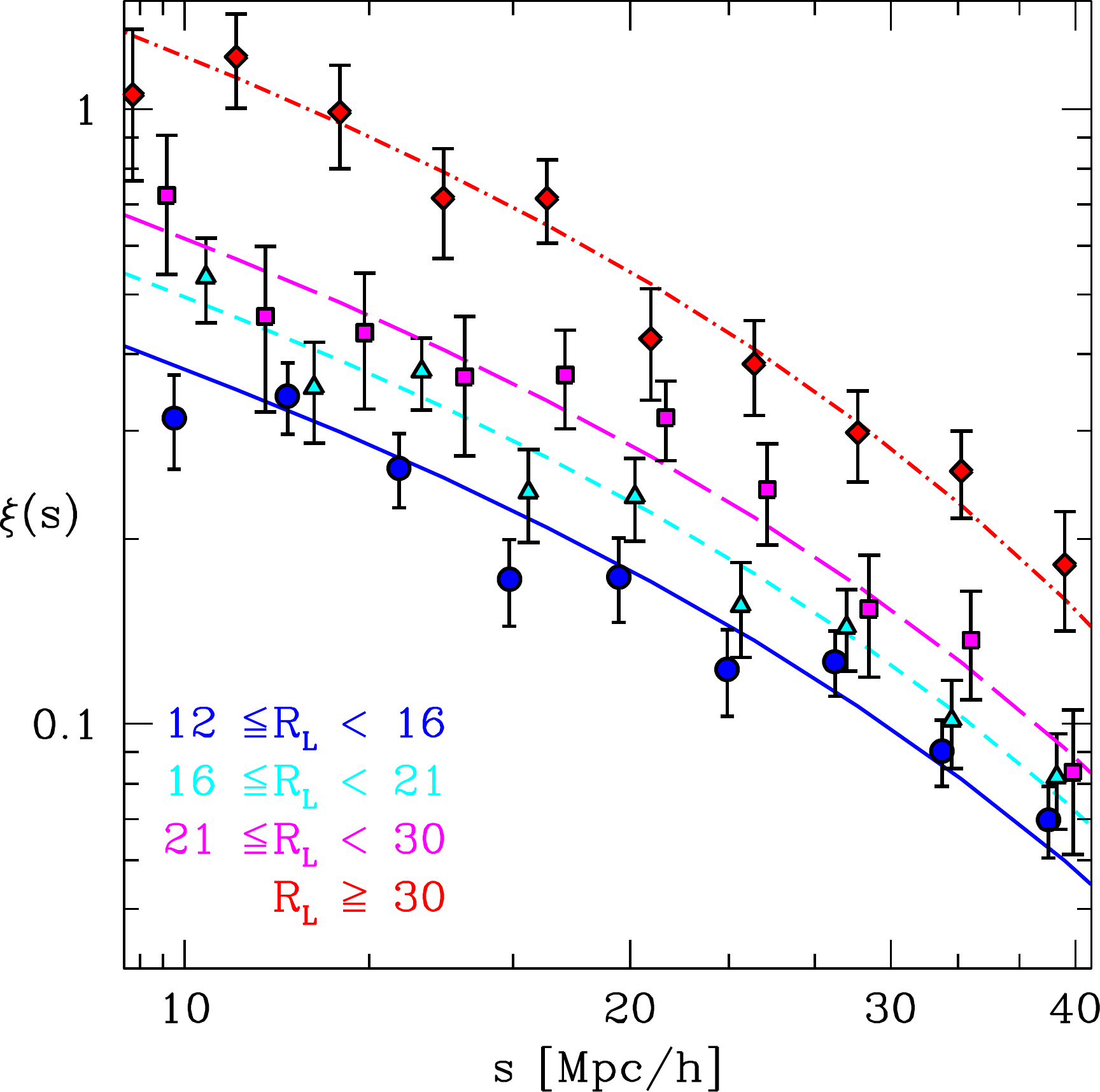}
   \caption{The redshift-space two-point correlation function of the four richness-selected cluster samples (dots), compared to the best-fit model obtained with Eq.~(\ref{e|genxi}, lines). The blue, magenta, purple, and red colour codes refer to the $12\le R_{L_*}<16$, $16\le R_{L_*}<21$, $21\le R_{L_*}<30$, and $R_{L_*}>30$, respectively. The error bars show the square roots of the diagonal elements of the covariance matrix.} 
   \label{f|xirl}
\end{figure}

The analysis to constrain the bias and $\sigma_8$ was performed with Monte Carlo Markov Chain technique, using the full covariance matrix. We adopted a standard likelihood, $\mathscr{L}_\mathrm{CL}\propto\exp(-\chi_\mathrm{CL}^2/2)$, with
\begin{equation}
    \chi_\mathrm{CL}^2 = \sum_{i=0}^{i=n} \sum_{j=0}^{j=n} (\xi_i - \hat{\xi}_i)
    \mathbfss{C}^{-1}_{ij} (\xi_j - \hat{\xi}_{j}) \, ,
    \label{e|chi2}
\end{equation} 
where $\xi_i$ is the correlation function measured in the $i$-th spatial bin, $\hat{\xi}_i$ is the model and $\mathbfss{C}^{-1}$ is the inverted covariance matrix. A  $\chi_\mathrm{CL}^2$ can be constructed for each richness sample.

As described in \S \ref{ss|rscf}, we fitted the redshift-space two-point correlation function $\xi(s)$ with the model given by Eq.~(\ref{e|genxi}). When the cosmology density parameters are fixed, as in our case, this model depends on three parameters only, the amplitude $\sigma_8$, the bias, $b$, and the cut-off scale $\sigma$, related to $\sigma_z$ through Eq.~(\ref{e|sigma}). Formally, the photo-$z$ error is a parameter to be determined too but, following the procedure described in \S \ref{ss|sze}, we assumed an informative Gaussian prior on $\sigma_z$, with a measured standard deviation of 0.003.

The redshift-space two-point correlation function of our photometric cluster sample is shown in Fig.~\ref {f|xi} (black dots). The error bars are the square root of the diagonal values of the covariance matrix. The dashed blue line shows the best-fit model given by Eq.~(\ref{e|genxi}). The red line is the result obtained by fitting only the dominant $\xi'$ term. The blue and red lines are in close agreement. The model with only $\xi'$ can fit the data nearly as well as the full model but it produces systematically shifted parameter estimates. If we neglect the Kaiser term, $f\mu^2$, we find a 6 per cent higher bias.

The long-dashed green line, that shows the case of a model without the photo-$z$ dumping term, clearly demonstrates the dramatic effect of photo-$z$ errors on the clustering shape.

Constraints are strongly degenerate. Clustering is strongly dependent on the product $b\ \sigma_8$. This degeneracy would be exact if the correlation function was measured in the real space. For the whole sample, we measured $b\ \sigma_8=2.29\pm0.08$, at the median redshift $z=0.37$. The error estimate is conservative, due to the choice of the prior standard deviation. Lowering the value of the $\sigma_z$ standard deviation enhances the precision of the measurement.

Figure~\ref{f|xirl} shows the redshift-space clustering measured in the four richness-selected samples. The amplitude of the correlation function scales according to the richness. The lines are the best-fit full models given by Eq.~(\ref{e|genxi}). The results are summarised in Table~\ref{tab_bias_sigma8}.

\section {Joint analysis}
\label{sec_join}

\begin{figure}
      \resizebox{\hsize}{!}{\includegraphics{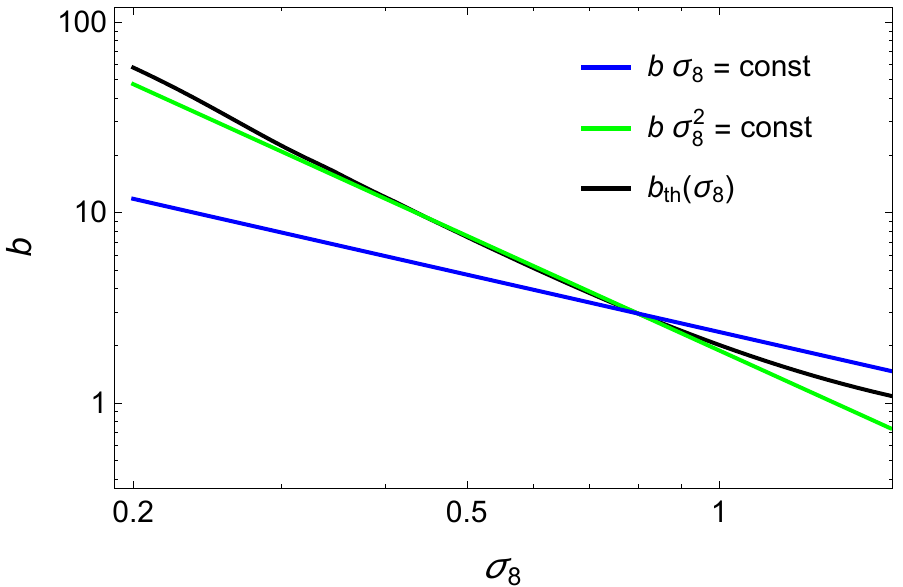}}
       \caption{Main degeneracies between bias and $\sigma_8$ as probed by either clustering (blue line), lensing (green line), or a theoretical modelling of the bias (black line). Theoretical predictions are based on \citet{tin+al10}. We considered $\sigma_8=0.8$ and the bias of a halo with mass $M_{200}=10^{14}M_\odot/h$ at $z=0.3$.}
	\label{fig_sigma8_bias_deg}
\end{figure}

\begin{table*}
\caption{Results from the stacked lensing (cols. 2, 3, and 4) and the clustering (cols. 5, 6, and 7) analysis of different binning in optical richness $R_{L^*}$ (col.~1). For the lensing part: number of clusters $N_\mathrm{cl}$ in the bin (col. 2); estimated cluster mass $M_{200}$ (col.~3); product $b\ \sigma_8^2$ (col.~4). For the clustering: number of clusters $N_\mathrm{cl}$ in the bin (col.~5); product $b\ \sigma_8$ (col.~6); estimated redshift error $\sigma_z$ (col.~7). Bias $b$ and $\sigma_8$ from the combined analysis are reported in cols. 8 and 9, respectively. Reported values of central estimate and dispersion are the bi-weight estimators of the posterior probability densities. Masses are in units of $10^{14}M_\odot/h$.}
\label{tab_bias_sigma8}
\centering
\begin{tabular}[c]{l  r r@{$\,\pm\,$}l  r@{$\,\pm\,$}l  r r@{$\,\pm\,$}l r@{$\,\pm\,$}l r@{$\,\pm\,$}l r@{$\,\pm\,$}l}
        \hline
	& \multicolumn{5}{c}{Lensing} & \multicolumn{5}{c}{Clustering} &  \multicolumn{4}{c}{Combined} \\
	& $N_\mathrm{cl}$	& \multicolumn{2}{c}{$M_{200}$} & \multicolumn{2}{c}{$b~\sigma_8^2$} &  $N_\mathrm{cl}$ & \multicolumn{2}{c}{$b~\sigma_8$} & \multicolumn{2}{c}{$\sigma_z$}	& \multicolumn{2}{c}{$\sigma_8$}		& \multicolumn{2}{c}{$b$}  \\
	\hline
	$R_{L^*} \ge 12$			&	1176		&	0.69&	0.11&	1.56	&	0.35	&	69527 	&	2.29	&	0.08 &	0.015 & 0.003	&	0.79		&	0.16	&	2.86	&	0.78	\\
	\hline
	$12\le R_{L^*} < 16$		&	476		&	0.48&	0.09&	1.87	&	0.46	&	29130	&	1.87	&	0.08	&	0.017 & 0.003	&	0.69	&	0.15	&	2.74	&	0.64	\\
	$16\le R_{L^*} < 21$		&	347		&	0.48&	0.10&	1.43	&	0.57	&	21047	&	2.12	&	0.08& 	0.016 & 0.003 	&	\multicolumn{2}{c}{}&	3.08	&	0.72	\\
	$21\le R_{L^*} < 30$		&	216		&	0.79&	0.12&	2.02 &	0.68	&	11962	&	2.30	&	0.16	&	0.015 & 0.004 	&	\multicolumn{2}{c}{}&	3.36	&	0.82	\\
	$30\le R_{L^*} $			&	137		&	1.92&	0.23&	1.86	&	0.83	&	7388		&	2.97	&	0.16& 	0.012 & 0.002	 &	\multicolumn{2}{c}{}&	4.31	&	1.03	\\
	\hline
	\end{tabular}
\end{table*}

In this section, we combine lensing and clustering to infer $\sigma_8$ and the halo bias. At the large scales probed by clusters of galaxies, we can safely consider the bias as linear. For each binning in optical richness, we can measure the weighted bias 
\beq
\label{eq_bias}
b(M_\mathrm{eff},z)=\int b(M_{200},z) f_\mathrm{sel}(M_{200}) d M_{200}, 
\eeq 
where $f_\mathrm{sel}(M_{200})$ is the selection function, 
\beq 
f_\mathrm{sel}(M_{200}) = \int_{R_{L^*}^\mathrm{min}}^{R_{L^*}^\mathrm{max}} P(M_{200}|R_{L^*}) n (R_{L^*})d R_{L^*}, 
\eeq 
$n (R_{L^*})$ is the distribution of the observed richness, $R_{L^*}^\mathrm{min}$ and  $R_{L^*}^\mathrm{max}$ delimit the richness bin, and $P(M_{200}|R_{L^*})$ embodies the mass-richness scaling relation through the conditional probability that a cluster with richness $R_{L^*}$ has a mass $M_{200}$. 

The big advantage of using the same clusters to measure lensing and clustering is that we do not need to model the bias to infer the amplitude $\sigma_8$. We are assured that we observe the same weighted bias, which is written in terms of an effective mass $M_\mathrm{eff}$, for both lensing and clustering. We do not need to determine the effective mass to estimate $\sigma_8$, even though $M_\mathrm{eff}$ can be constrained with the lensing fit of the central halos.

This basic approach does not need any derivation of the scaling relation between the observable (the optical richness in our case) and the cluster mass. The effects of the cluster selection function and of the scaling relation are included in the effective bias. In this way, we avoid two of the main difficulties which plague cosmological tests based on cluster of galaxies \citep{se+et14,ser+al14b}.

Bias and $\sigma_8$ can be computed by properly matching the constraints obtained with either lensing (which is degenerate with $b\,\sigma_8^2$) and clustering (which is degenerate with $b\,\sigma_8$). Degeneracies between bias and $\sigma_8$ which affect each of the two probes are pictured in Fig.~\ref{fig_sigma8_bias_deg}. Results are summarised in Table~\ref{tab_bias_sigma8}. We explored two methods to infer $\sigma_8$ according to whether or not the information on the mass halo, which is inferred from the small scales in the lensing analysis, is used to infer the amplitude of the power spectrum, $\sigma_8$.

\subsection{First method}

\begin{figure}
      \resizebox{\hsize}{!}{\includegraphics{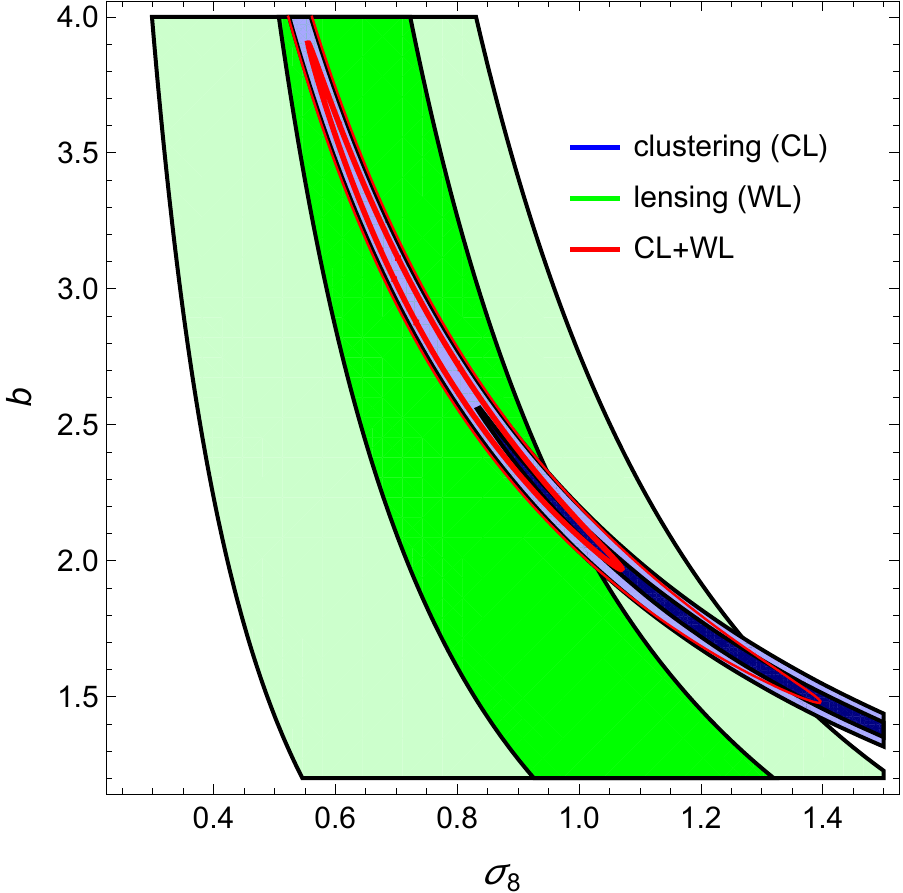}}
       \caption{Derived probability density of bias $b$ and $\sigma_8$ in the reference case ($R_{L_*}\ge 12$). The green (blue) regions include the confidence regions as obtained from lensing (clustering). The darker (lighter) area includes the 1-(3-)$\sigma$ confidence region in two dimensions, here defined as the region within which the value of the probability is larger than $\exp(-2.3/2)$ ($\exp(-11.8/2)$) of the maximum. The red thick (thin) contour includes the 1-(3-)$\sigma$ confidence regions from the joint analysis.}
	\label{fig_sigma8_bias}
\end{figure}

\begin{figure}
      \resizebox{\hsize}{!}{\includegraphics{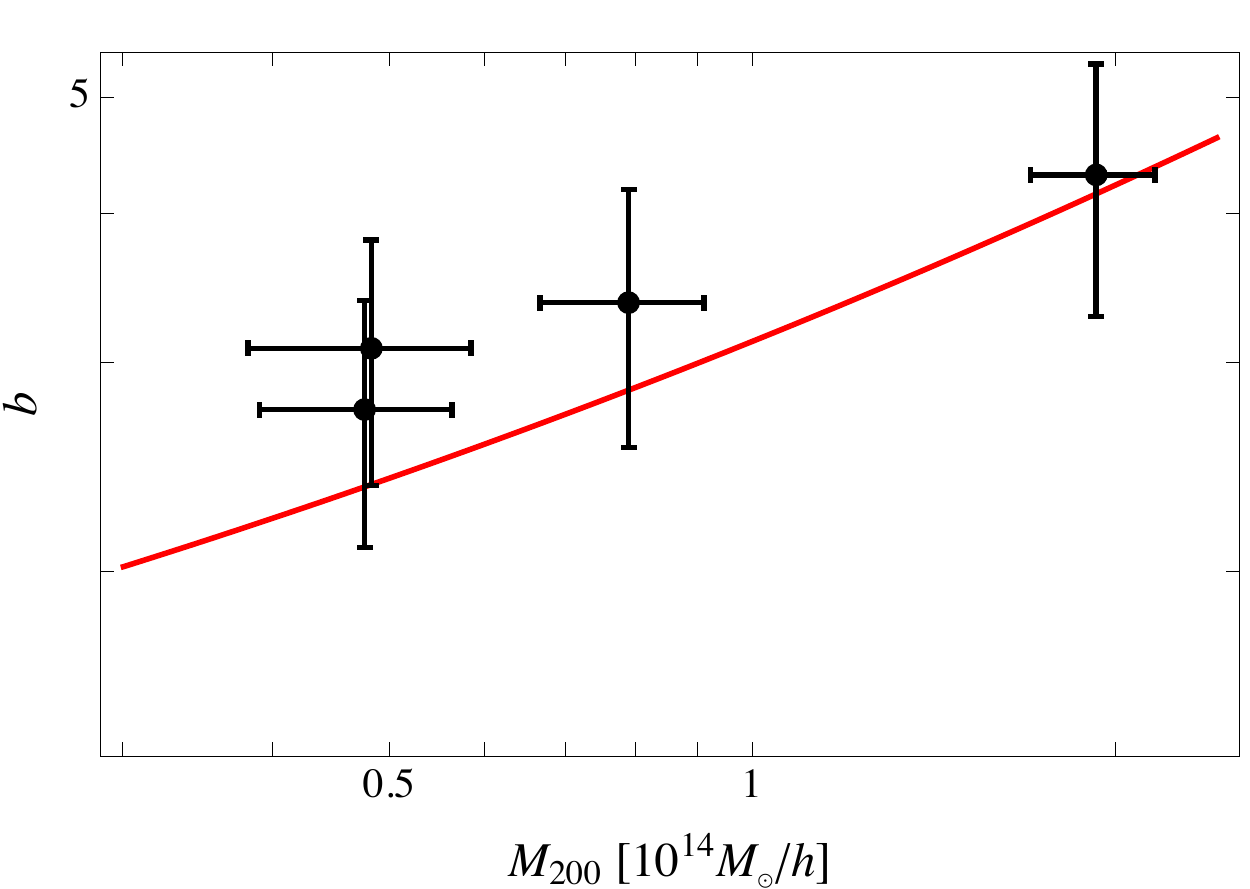}}
       \caption{Bias as a function of mass. The red line is the prediction by \citet{tin+al10} for $\sigma_8=0.8$ at $z=0.37$.}
	\label{fig_bias_M200}
\end{figure}

The first proposed method is the most basic. It consists in a simultaneous analysis of $\sigma_8$ and bias. It uses only the cosmological information derived from the 2-halo term in the lensing modelling. Together with $\sigma_8$, we could then determine the halo bias $b_i$ for each binning in optical richness. This method fully exploits the fact that we are measuring clustering and lensing for the same clusters, which were selected based on their richness. The determination of $\sigma_8$ does not require the modelling of the halo bias as a function of mass/richness.

The joint analysis was performed with the combined likelihood
\beq
\mathscr{L}_\mathrm{tot}\left( \sigma_8,\{ b_i\} \right) \propto \prod_i \mathscr{L}_{\mathrm{GL},i} \mathscr{L}_{\mathrm{CL},i},
\eeq
where $\mathscr{L}_{\mathrm{CL},i}$ is the clustering likelihood for the $i$-th richness bin, see Eq.~(\ref{e|chi2}). The dependence on $\sigma_z$ was marginalised over. The lensing likelihood, $\mathscr{L}_{\mathrm{GL},i}(b_i\ \sigma_8)$, was obtained by marginalising the posteriori probability distribution obtained with the stacked lensing analysis of the $i$-th bin, see Sec.~\ref{sec_lens_cons}. We adopted uniform priors for $\sigma_8$ and the bias. Results for the reference case are reported in Table~\ref{tab_bias_sigma8}. We found $\sigma_8=0.79\pm0.16$ and a bias of $2.86\pm0.78$. Confidence regions are plotted as red contours in Fig.~\ref{fig_sigma8_bias}. Even though the lensing constraints are quite shallow with respect to the clustering results, they are crucial to break the degeneracy between bias and power spectrum amplitude. 

The lensing constraints are exactly degenerate with the contours where $b\ \sigma_8^2$ is constant whereas the constraints from clustering align to a very good approximation with the loci of points where $b\ \sigma_8$ is constant. The latter degeneracy would be exact if the cluster-cluster correlation function was measured in the real space. A simplified joint likelihood can be written in terms of the $\chi^2$-function
\beq
\chi^2 = \sum_i \left( \frac{\left[b\ \sigma_8^2\right]_i-b_i  \sigma_8^2}{\delta \left[b\ \sigma_8^2\right]_i} \right)^2
+\left( \frac{\left[b\ \sigma_8\right]_i-b_i  \sigma_8}{\delta \left[b\ \sigma_8\right]_i} \right)^2,
\eeq
where the sum runs over the different richness bins; $\left[b\ \sigma_8^2\right]_i$ and $\left[b\ \sigma_8\right]_i$ are the measurements from lensing and clustering in the $i$-th bin, respectively. For the following tests, we used the simplified version of the likelihood.

We first checked for consistency. Estimates of $\sigma_8$ obtained considering subsamples in optical richness one at a time are in agreement among themselves and with the reference case. We got $\sigma_8=0.92\pm0.27$ for $12 \le R_{L^*} < 16$, $\sigma_8=0.59\pm0.24$ for $16 \le R_{L^*} < 21$, $\sigma_8=0.80\pm0.29$ for $21 \le R_{L^*} < 30$ and $\sigma_8=0.57\pm0.22$ for $R_{L^*} \ge 30$.

Being the estimates consistent, we could analyse the four subsamples together. In this way we could measure at the same time the halo bias in each richness bin and the amplitude of the power spectrum. We found $\sigma_8=0.69 \pm 0.15$, which is fully consistent with the reference case. 

The measured halo bias is an increasing function of the optical richness, see Table~\ref{tab_bias_sigma8}. Stacked lensing also provided direct estimates of effective masses thanks to the modelling of the main halo term. We could then look for trends of $b$ with mass without assuming any scaling relation. The bias increases with mass in agreement with results from theoretical predictions, see Fig.~\ref{fig_bias_M200}.

We remark that we used the small scale regime only to derive the halo mass whereas we did not try to extract constraints on the bias from the regions within the viral radius. The determination of $\sigma_8$ and of the bias for each bin was independent of the small scale-regime, which entered only when we studied the evolution of bias with mass. In this scheme the effective mass is identified with the lensing mass, which is an acceptable approximation for stacking analyses in physical length units \citep{oka+al13,ume+al14}.

\subsection{Second method}

In the second approach, we focused on the determination of $\sigma_8$ by assuming that the bias is a known cosmological function of the peak height \citep{tin+al10,bha+al11}. This requires the knowledge of the halo mass, which is determined with the stacked lensing. The weight factor when stacking in physical length units is mass-independent and estimated masses and concentration are not biased \citep{oka+al13,ume+al14}. We then assumed that the effective mass measured by lensing for the central halo is the same effective mass probed by the bias, see Eq.~(\ref{eq_bias}), which  strictly holds if the signals are linear in mass. The quantitative analysis can be performed in terms of a $\chi^2$ function,
\begin{multline}
\label{eq_chi_sigma8_th}
\chi^2  \left( \sigma_8,\{ M_{200,i}\} \right)  =  \sum_i  \left[ \frac{\left( \left[b\,\sigma_8^2\right]_i-b_{\mathrm{th},i}  \sigma_8^2 \right)^2}{\delta \left[b\ \sigma_8^2\right]_i^2 + \delta \left[b_{\mathrm{th},i}\right]^2  \sigma_8^4} \right. \\
+  \left. \frac{\left( \left[b\ \sigma_8\right]_i-b_{\mathrm{th},i}  \sigma_8 \right)^2}{\delta \left[b\ \sigma_8\right]_i^2 + \delta \left[b_{\mathrm{th},i}\right]^2  \sigma_8^2} 
 + \left( \frac{M_{200,i}-M_{200,i}^\mathrm{obs}}{\delta M_{200,i}^\mathrm{obs} } \right)^2 \right].
\end{multline}
where $b_{\mathrm{th},i}=b_{\mathrm{th},i}(\sigma_8, z_i, M_{200,i})$ is the theoretical prediction for given $\sigma_8$ and halo mass $M_{200,i}$ at redshift $z_i$ and $\delta \left[b_{\mathrm{th},i}\right]$ is the related uncertainty. We used the fitting formula for the bias derived in \citet{tin+al10}. They found a six per cent scatter about their best-fit relation, which we conservatively adopted as the uncertainty on the theoretical prediction.

Differently from the first approach, where the biases themselves were free parameters, now they are expressed in terms of $M_{200}$. Since the masses were already constrained by the lensing analysis, we added to the $\chi^2$ function a penalty term, i.e., the third right hand term in Eq.~(\ref{eq_chi_sigma8_th}), and we still formally considered the mass associated to each bin as a parameter to be determined. Of course, the posterior estimate of each $M_{{200},i}$ just follows the prior but we had to include the penalty not to underestimate the error on $\sigma_8$. As for the first approach, the second method still does not need to calibrate the mass-richness relation.

For the reference binning in optical richness ($R_{L^*} \ge 12$), we obtained $\sigma_8=0.75\pm0.08$, in agreement with what obtained with the first approach. The use of the information on the dependence of the bias on the peak height nearly halved the statistical error. The theoretical constraint on the bias is nearly degenerate with the lensing one, i.e., $b_\mathrm{th}$ is nearly proportional to the inverse squared $\sigma_8$. Constraints from clustering, lensing or theoretical predictions are compared in Fig.~\ref{fig_sigma8_bias_deg}. However, we can not use the theoretical constraint without the information on mass from lensing.

\section{Forecasting}
\label{sec_fore}

The accuracy in the determination of $\sigma_8$ will greatly benefit from future optical galaxy surveys. As a test bed, we consider the wide survey planned by the Euclid mission\footnote{\url{http://www.euclid-ec.org/}}. The signals of either lensing or clustering can be enhanced by considering a larger number of clusters (which can be achieved with either deeper or wider surveys), a larger number of background sources (deeper surveys) and a larger survey area in order to cover the lensing 2-halo term up to 50~Mpc (wider surveys). With regard to these three aspects, Euclid will represent a significant improvement with respect to the data-sets we considered in the present paper.

Euclid will observe an area of $15000~\deg^2$ and it is expected to detect $n_\mathrm{g}\sim 30$ galaxies per square arcminute with a median redshift greater than 0.9, that can be used for weak lensing analyses \citep{eucl_lau_11}. These basic properties are enough to forecast the expected accuracy in the $\sigma_8$ determination from the joint lensing plus clustering analysis we presented.

The area of the Euclid survey is nearly 2 times larger than the area we considered in the clustering analysis and nearly 100 times wider than the CFHTLenS, with a corresponding expected improvements in the corresponding signals. 

Due to improved photometry, a larger number of clusters will be detected to higher redshifts. Most of the newly detected clusters will be low mass halos producing a small lensing signal. On the other hand, Euclid will significantly extend the redshift range of the background galaxies, whose lensing signal is maximised at high redshift.

Recently, \citet{for+al14b} presented the CFHTLenS 3D-Matched-Filter catalog of cluster galaxies. Candidate clusters where selected if they had at least two member galaxies within the virial radius and a detection significance in excess of $3.5$. More than $N_\mathrm{cl}\sim 18000$ clusters were detected in the $\sim150\deg^2$ area of the survey in the redshift range $0.2\ls z\ls0.9$. More than 14000 candidate clusters had an estimated $N_{200}>10$. By comparison, with SDSS-III quality data, \citet{wen+al12} detected $N_\mathrm{cl}\sim 1200$ clusters with $N_{200}\gs 8$ in the redshift range $0.1 <z< 0.6$ over the area of $\sim 130\deg^3$ in common with the CFHTLenS. We can conclude than nearly 10 times more clusters can be identified by increasing the photometric depth of the survey from SDSS-III to CFHTLenS quality data. 

The Euclid mission is expected to identify an even larger number of clusters. Nevertheless, a significant number of them will be made of small groups, whose photometric redshift determination might be uncertain, which hampers the clustering analysis. Furthermore, the larger the number of identified clusters, the larger their density in the sky. As we have seen, the shot noise is not the only source of uncertainty at large radii. If we consider clusters  whose mean separation is smaller than the range over which we measure the shear profile, we can not simply rescale the lensing signal-to-noise ratio as SNR $\propto \sqrt{N_\mathrm{cl}}$. We can then conservatively consider an improvement of a factor $\sim10$ in the clustering/lensing signal detected by Euclid with respect to the present analysis due to the larger density of detected clusters.

The background lensed galaxies resolved by Euclid will be more and further away than the sources in the CFHTLenS. The number density of galaxies in the CFHTLenS with shear and redshift data is $n_\mathrm{gal}\sim17$ galaxies per square arcminute \citep{hey+al12}. The effective weighted galaxy number density that is useful for a lensing analysis is $n_\mathrm{gal}\sim11$ galaxies per square arcminute in the redshift range $0.2<z<1.3$ with a median redshift of $z\sim 0.75$ \citep{hey+al12}, and it is $\sim 6$ galaxies per square arcminute at $z\ga 0.7$ \citep{cov+al14}. Euclid sources will be more numerous ($n_\mathrm{g}\sim 30$ galaxies per square arcminute) and at a median redshift of $\sim0.9$. These two factors make the signal behind a lens nearly two times larger.

Finally, the wide coverage of the Euclid survey will enable us to detect the 2-halo lensing signal to its full radial extent. The extension of the radial coverage from 15 up to $30~\mathrm{Mpc}/h$ can decrease the errors in the lensing estimate of $b\ \sigma_8^2$ from $\gs 20$ to $\sim 15$ per cent, see Table~\ref{tab_bias_sigma8Sq}. The improvement would be even more significant considering radii up to $50~\mathrm{Mpc}/h$ but this effect would be counterbalanced by the increased overlap in the lensing area of near clusters. 

Based on the above considerations, we expect that stacked lensing can measure $b\ \sigma_8^2$ with an accuracy of $\sim 0.3$ per cent and that clustering can measure $b\ \sigma_8$ with an accuracy of $\sim 0.1$ per cent. The combined effects of a larger sample of clusters and less noisy measurements should be enough to get an accuracy $\delta \sigma_8 \sim 0.003$ with the Euclid mission without any assumption on the mass-richness scaling or any modelling of the halo bias. 

In the present analysis we could keep $\Omega_\mathrm{M}$ fixed because of the large statistical uncertainties. This will be no more the case in presence of Euclid quality data. On one hand, the dependence on $\Omega_\mathrm{M}$ of the joint clustering plus lensing analysis enlarge the forecasted statistical uncertainty on $\sigma_8$. On the other hand, $\Omega_\mathrm{M}$ could be determined to very good accuracy by exploiting other features of the joint analysis that we discuss in the following section as well as by combining the present method with other external probes. These effects should counterbalance each other and the expected $\delta \sigma_8$ should be nearly unchanged.

The estimate of $\sigma_8$ will also greatly benefit from the increased spectroscopic sample associated to the Euclid survey. Working with spectroscopic rather than photometric redshift eliminates one of the main sources of uncertainty in the clustering analysis, see Sec.~\ref{ss|sze}.

The high precision in Euclid measurements will demand for very accurate theoretical modelling. The bias is a stochastic process which is difficult to model. In this sense our basic approach, where the bias is treated as an effective parameter, is very promising.

\section{Prospects}
\label{sec_pros}

In this section, we consider some improvements that might enhance the performances of joint analyses of lensing and clustering.

\subsection{Cosmological dependences}

Due to the limited field of view covered by the lensing catalogue, we did not aim at constraining all cosmological parameters. However, the joint analysis of cross-correlation makes it possible to constrain cosmological density parameters too.

The lensing signal is proportional to the matter density $\bar{\rho}_\mathrm{m}$, see Eqs.~(\ref{eq_Sigma},~\ref{eq_Delta_Sigma}). On fully linear scales, stacked lensing and clustering can then constrain the product $\sigma_8 \Omega_\mathrm{M}$. Neglecting neutrinos we can single out the linear growth factor $D_+$ at the lens redshift too, so that the inferred product becomes $\sigma_8 \Omega_\mathrm{M}D_+(\Omega_\mathrm{M}, \Omega_\mathrm{\Lambda},...)$. Cosmological parameters enter with a lesser impact also in the angular diameter distances and the shape of the power spectrum.

The inference of cosmological densities requires that the measured signals are independent of the reference cosmological framework. Stacking in physical length and the estimation of the lensing signal as done in this paper require the assumption of a cosmological model. In fact, angular diameter distances have to be computed to convert angular separations to physical lengths, to convert tangential shears to surface density contrast, and to weight the lensing signal for sources at different redshifts. These effects are usually considered small and neglected. The stacked lensing signal derived in a fixed cosmology has been already used to constrain the cosmological density parameter $\Omega_\mathrm{M}$ \citep{man+al13,cac+al13}.

An unbiased procedure would require stacking the signal in angular annuli for lenses in small redshifts bins and adopting a single population of background source galaxies for the lensing analysis \citep{og+ta11}. This approach is suitable for large surveys, like Euclid.

\subsection{Self-calibration}

The approach we presented circumvents the problems inherent to a proper calibration of the cluster mass-observable relation. Cluster properties which can be easily measured for a large number of objects in ongoing and future large surveys, such as optical richness and X-ray luminosity, can be used as mass proxies. This requires the accurate calibration of the observable through comparison with direct mass estimates such as weak lensing determinations of massive clusters or X-ray analyses assuming hydrostatic equilibrium \citep{ras+al12}. However, these estimates are scattered too, with intrinsic scatters of $\sim$15 per cent for lensing masses, and of $\sim$25 per cent for X-ray masses \citep{se+et14}. The level of bias is more difficult to ascertain because of differences as large as $40$ per cent in either lensing or X-ray mass estimates reported by different groups \citep{se+et14}.

In principle, the combination of the two observables we considered in this paper, i.e, the stacked lensing and the clustering, enables the self-calibration of the mass-observable relation together with the calibration of the other major source of systematic errors, i.e., the photometric redshift uncertainty \citep{og+ta11}. The degeneracy between these two uncertainties can be broken by observing tangential shear signals over a wide range of radii \citep{og+ta11}. Accurate self-calibration of systematic errors can be indeed attained in future surveys, when we expect a redshift accuracy in lensing tomographic bins of $\sim$0.1 per cent and that the mean cluster mass in each bin can be calibrated to 0.05 per cent \citep{og+ta11}.

\section{Conclusions}
\label{sec_conc}

Haloes are biased tracers of the underlying matter distribution. The bias is mass dependent, and, for a given mass range, it is a non-linear and stochastic function of the underlying matter density field. In the CDM scenario for structure formation and evolution, the knowledge of the relative abundance of haloes is sufficient to approximate the large-scale bias relation \citep{sh+to99}. Detailed knowledge of the merger histories of dark matter haloes is not required. The peak-background split achieves an agreement at the level of $\sim 20$ per cent with the numerical results, both at high and low masses \citep{tin+al10}.

The knowledge of the halo bias plays an important role in the determination of cosmological parameters through tests based on the abundance and clustering of high-mass haloes \citep{tin+al10,og+ta11}. Methods that utilise correlation functions to constrain cosmology require precise knowledge of halo clustering, which is understood in terms of the bias of the haloes in which structures collapse. Furthermore, the information obtained from the bias of clusters is complementary to their abundance. Indeed, proper self-calibration of cluster surveys relies on the additional information present in clustering data.

The analysis of stacked lensing and clustering can directly measure the bias as a function of mass and test this important prediction of the $\Lambda$CDM paradigm. Combining results from the two-point correlation function of a photometric sample of 69527 clusters selected from the SDSS with the weak lensing analysis of 1176 clusters from the CFHTLenS, we measured the linear bias in the mass range $0.5 \la M_{200}/(10^{14}M_\odot/h) \la 2$. We found excellent agreement with results from dark matter $N$-body simulations \citep{tin+al10,bha+al11}. The bias scales with mass according to theoretical predictions. Thanks to the minimal modelling of the employed method, we could obtain this result bypassing the calibration of the scaling relation between cluster mass and observable.

The development of independent methods to measure cosmological parameters and a proper comparison of estimated values from different experiments is crucial to test possible failures of the standard $\Lambda$CDM model or hidden systematics. We could determine the power spectrum amplitude with an accuracy $\delta\sigma_8 \gs 0.1$. Even though we fixed $\Omega_\mathrm{M}=0.3$ in our analysis, the statistical error is too large to discriminate between the discrepant estimates of $\sigma_8$ from either number counts \citep{planck_2013_XX} or CMB \citep{planck_2013_XVI}. The method is nevertheless promising for its minimal modelling and well controlled systematics. Our estimation of $\sigma_8$ does not rely on any mass-observable scaling relation. In view of future wide surveys like e.g. Euclid \citep{euclid_ame+al13}, our method, as well as other proposed approaches which combine galaxy/matter correlation functions, provides complementary information to constrain cosmological parameters and should help to test non standard physics. Future deep galaxy surveys will enable to detect and stack lensing clusters in small redshifts bins up to $z\la 1.5$. This will enable to further break parameter degeneracies and limit systematics \citep{og+ta11}.

\section*{Acknowledgements}
The authors thanks C. Carbone for useful discussions. The authors acknowledge financial contributions from contracts ASI/INAF n.I/023/12/0 `Attivit\`a relative alla fase B2/C per la missione Euclid', PRIN MIUR 2010-2011 `The dark Universe and the cosmic evolution of baryons: from current surveys to Euclid', and PRIN INAF 2012 `The Universe in the box: multiscale simulations of cosmic structure'. This work is based on observations obtained with MegaPrime/MegaCam, a joint project of CFHT and CEA/IRFU, at the Canada-France-Hawaii Telescope (CFHT) which is operated by the National Research Council (NRC) of Canada, the Institut National des Sciences de l'Univers of the Centre National de la Recherche Scientifique (CNRS) of France, and the University of Hawaii. This research used the facilities of the Canadian Astronomy Data Centre operated by the NRC of Canada with the support of the Canadian Space Agency. CFHTLenS data processing was made possible thanks to significant computing support from the NSERC Research Tools and Instruments grant program.


\setlength{\bibhang}{2.0em}

\end{document}